\documentclass[journal = jctcce, layout=twocolumn, manuscript=article]{achemso}
\setkeys{acs}{articletitle = true, maxauthors = 6, super=true}
\usepackage[colorlinks, linkcolor=blue, anchorcolor=blue, citecolor=blue]{hyperref}
\usepackage[utf8]{inputenc}
\usepackage[T1]{fontenc}
\usepackage{amsmath} 
\usepackage{cleveref}  
\usepackage{multirow}
\usepackage{amssymb}
\usepackage{diagbox}
\usepackage{xcolor}
\usepackage{booktabs}
\usepackage{graphicx}
\usepackage{subcaption}
\usepackage{pdfpages}
\SectionNumbersOn  
\SectionsOn  
\AbstractOn  

\author{Zhenxing Zhu}
\affiliation[Beijing Normal University]
{Key Laboratory of Theoretical and Computational Photochemistry of Ministry of Education, College of Chemistry, Beijing Normal University, Beijing 100875,
China;}

\author{Diandong Tang}
\email{tangdd@uw.edu}
\affiliation[University of Washington]
{Department of Chemistry, University of Washington, Seattle, Washington 98195, USA;}

\author{Lin Shen}
\email{lshen@bnu.edu.cn}
\affiliation[]
{Key Laboratory of Theoretical and Computational Photochemistry of Ministry of Education, College of Chemistry, Beijing Normal University, Beijing 100875,
China;}

\author{Wei-Hai Fang}
\affiliation[Beijing Normal University]
{Key Laboratory of Theoretical and Computational Photochemistry of Ministry of Education, College of Chemistry, Beijing Normal University, Beijing 100875,
China;}

\title[Short running title]
{Fewest-Switches Surface Hopping with Combined Deep Learning Potential and Long Short-Term Memory Network Propagator for Simulating Realistic Photochemical Processes}
\begin{document}  
\maketitle
\begin{abstract}  
Fewest-switches surface hopping (FSSH) is the most popular method for simulating photochemical processes of molecular systems. Recently, we have constructed long short-term memory (LSTM) networks as a propagator for electronic subsystems in FSSH dynamics simulations. The collective results on Tully's three models have been reproduced satisfactorily. In the present work, we develop an extended LSTM-FSSH framework to simulate realistic photochemical reactions. The input features of LSTM as well as the training procedure are redesigned to represent high-dimensional nuclear degrees of freedom in an effective way. Equivariant neural networks are integrated with LSTM to build adiabatic potential energy surfaces in ground and excited states. Photoisomerizations of $\mathrm{CH_2NH}$ and azobenzene are simulated, showing that our new proposed LSTM-FSSH method can produce excited-state lifetimes and product yields accurately in comparison with conventional FSSH simulations as reference. Only 10 reference trajectories are required for training LSTM networks, and then a trajectory ensemble can be generated with very efficient LSTM-FSSH dynamics simulations to obtain collective results.
\end{abstract}  
\section{INTRODUCTION}
Classical molecular dynamics simulations cannot study photochemical reactions because the Born-Oppenheimer approximation is no longer applicable in these cases. Fully quantum treatment on both nuclear and electronic degrees of freedom is too computationally expensive for realistic molecular systems, if not impossible. Mixed quantum classical molecular dynamics (MQC-MD) method, including Ehrenfest mean-field (EMF), surface hopping and ab initio multiple spawning (AIMS), provides a powerful tool to chemists. One of the most popular MQC-MD algorithms is the fewest switches surface hopping (FSSH)\cite{Tully10.1063/1.459170,Tully10.1063/1.467455} developed by Tully in 1990. During FSSH dynamic simulations, the nuclei move classically on an adiabatic potential energy surface (PES) with a certain probability of jumping to any other state. The transition probability is obtained based on the time evolution of electron density matrix, which incorporates quantum feedback of electronic motion. Despite its simple implementation and great success in many applications, the computational expense of FSSH remains much higher than that of ground-state MD. Not only the potential energies and gradients for all relevant electronic states but also the nonadiabatic coupling vectors (NACVs) between different states should be calculated at a high level of electronic structure theory in each time step. Furthermore, FSSH requires an ensemble of trajectories to estimate collective results such as the excited-state lifetime and product yield. Strictly speaking, a dozen FSSH trajectories make little sense for revealing any mechanism of medium-sized or larger polyatomic molecules.

In recent years, there has been a growing interest in integrating machine learning (ML) techniques with MQC-MD for accelerating nonadiabatic dynamic simulations. ML has demonstrated its strong capability of fitting potential energy surfaces of high-dimensional molecules, either in the ground or excited state\cite{PhysRevLett.98.146401,WANG2018178,Pavlo:doi.org/10.1002/jcc.26004,doi:10.1021/acs.chemrev.0c00749,doi:10.1021/acs.chemrev.4c00572}. However, accurate predictions of NACVs are much more difficult because the value of NACV often remains to be zero in most regions of PES but changes very suddenly around conical intersections, exhibiting significant scarcity and singularity characteristic. This problem may become more serious with the growth of molecule size and involved electronic states. Despite some great progress such as the combination of SchNet and SHARC\cite{C9SC01742A,Marquetanddoi:10.1021/acs.jpclett.0c00527}, an alternative ML framework without the need of fitting NACVs is always attractive to the field of theoretical and computational photochemistry. 

Cui, Lan, Wen and their co-workers have independently implemented the Zhu-Nakamura molecular dynamics \cite{Zhu10.1063/1.467877,Zhu10.1063/1.469057} simulation on ML-based adiabatic PESs\cite{doi:10.1021/acs.jpclett.8b03026,Landoi:10.1021/acs.jpclett.8b00684,Wenjingdoi:10.1021/acs.jpca.3c01036,WenjingD4CP01497A} 
{
based on improvement of Landau-Zener formalism. Unlike FSSH, the transition probability can be estimated based only on adiabatic energies and gradients. However, the difference between the collective results using FSSH and Zhu-Nakamura methods should be examined carefully when studying a new photochemical reaction.
Barbatti et. al. has developed a novel model for approximating nonadiabatic couplings based on time-dependent Baeck-An (TD-BA) approximation with only energy gaps and its second time derivative.\cite{An17_064107,Barbatti22_13624}
Truhlar et. al. developed generalized semiclassical Ehrenfest method combining curvature-driven approximated nonadiabatic coupling and gradient correction on time-derivative matrix (TDM) scheme for momentum conservation. Their method can be easily extended into intersystem crossing domain with spin-orbit coupling.
\cite{Truhlar22_1320,Truhlar24_4396}
}
Rapid development on deep neural networks (NNs), such as long short-term memory (LSTM)\cite{6795963} and transformer\cite{vaswani2023attentionneed}, provides an opportunity to accelerate dynamic simulations by replacing the equation of motion with ML-predicted sequences. For time propagation of electronic degree of freedom, the history of density matrix can be considered as a time series and applied to LSTM or other NNs to predict the electron density matrix in the next time step. {For example, Lan and coworkers developed several LSTM models and successfully reproduced dynamics} simulation results of the symmetrical quasi-classical dynamics method based on the Meyer-Miller mapping Hamiltonian (MM-SQC)\cite{Landoi:10.1021/acs.jpclett.2c02159} and the full quantum multilayer multiconfiguration time-dependent Hartree (ML-MCTDH) method\cite{doi:10.1021/acs.jpclett.1c02672}. Singh et al. employed Fourier neural operators as an effective time propagator of the MCTDH quantum dynamics for time-independent and time-dependent potentials\cite{https://doi.org/10.1002/jcc.27443}. Lin and Gao integrated convolutional neural networks and LSTM for prediction of open quantum dynamics under non-Markovian stochastic Schrödinger equation\cite{doi:10.1021/acs.jctc.5c00215}. Ullah et al. proposed physics-informed and Lie algebra-based approaches to recover exact trace conservation during ML-driven quantum dynamics\cite{10.1063/5.0266604}. Sun and coworkers built different kinds of NNs and investigated the influence of neural network complexity on time evolution of electronic state population\cite{Sun10.1063/5.0073689}. Recently, our group implemented LSTM as a propagator for electronic subsystem during the FSSH simulations\cite{doi:10.1021/acs.jpclett.2c02299}. The LSTM-FSSH satisfactorily reproduced collective results on Tully’s three models.

Although the applications of ML-based propagators have covered from MM-SQC to ML-MCTDH, the study on realistic photochemical reactions is rare now. In comparison with other nonadiabatic dynamics methods, surface hopping is more suitable for simulating complex molecular systems. However, there are still two difficulties that impede further application of our constructed LSTM-FSSH framework. First, the dimensionality of nuclear subsystem varies in a broad range and sometimes significantly exceeds that of electronic subsystem. { How to perform the modeling of the coupling between nuclear and electronic motions is not only the core of MQC-MD but also the key to the performance of ML prediction.} The information about two subsystems should be both involved in LSTM as input features, but the balance point between the amounts of nuclear and electronic information is still being explored case by case. Second, the excited-state calculation on potential energies and gradients is a bottleneck for most photochemical reaction systems. The combination of ML-based force fields and LSTM-FSSH provides a solution. However, the construction of ML-based potentials in excited states is much more challenging than that in the ground state, mainly due to the lack of reference values and the complexity of intersection regions on PESs.

In this work, we aim to solve the above problems and develop an extended LSTM-FSSH framework to simulate realistic photochemical reactions more effectively. The input features of LSTM were redesigned to represent high-dimensional nuclear degrees of freedom, and the procedure was modified to enhance training efficiency. The LSTM was further integrated with the NequIP model to build adiabatic PESs. The method will be described in Section \ref{sec:METHOD}. Then two photoisomerization reaction systems were employed to validate the performance of LSTM-FSSH. The results will be provided and compared with conventional FSSH simulations in Section \ref{sec:RESULTSANDDISC}, followed by conclusion and outlook in Section \ref{sec:CONCLUSIONSANDOUTLOOK}.

\section{METHOD}  

\label{sec:METHOD}
\subsection{Equivariant neural network}
Conventional deep neural networks have been widely used to learn potential energy surfaces of diverse molecules\cite{NIPS2017_303ed4c6,10.1063/1.5019779}, including photochemical reaction systems. Similarly, NequIP\cite{Batzner2022} also employs atomic types and coordinates as inputs, and then outputs molecular potential energies and gradients. The distinctive feature of NequIP lies in achieving equivariance through the utilization of equivariant convolutional layers, which aims at attaining corresponding alterations in the outcomes under symmetric operations. For instance, in the steerable E(3) equivariant graph neural networks, the forces on atoms are predicted directly\cite{brandstetter2022geometricphysicalquantitiesimprove}. If the molecular system rotates, the predicted atomic forces should also rotate accordingly, which represents equivariance. The strategy applied to NequIP is to encode the atomic positions and other feature information by means of spherical harmonic functions, followed by equivariant convolutional layers for information interaction to guarantee the realization of equivariance in the transmission of network information. A growing number of applications in various systems have demonstrated its superior accuracy, especially in absence of big data for training. Here we built NequIP to predict PESs of photochemical systems based on the e3nn library that merges recent progress on neural network models, such as tensor field networks\cite{thomas2018tensorfieldnetworksrotation}, 3D steerable convolutional neural networks\cite{weiler20183dsteerablecnnslearning}, and Clebsch-Gordan Nets\cite{kondor2018clebschgordannetsfullyfourier}. More details can be referred to the original literature.

\subsection{FSSH method}
In the FSSH method, the atomic nuclei move on a single potential energy surface in an adiabatic electronic state, which obeys the Newton's second law as
\begin{align}
    m_a {\boldsymbol{\ddot{R}}}_a &=-\boldsymbol{\nabla_{R_a} } E_j (\boldsymbol{R}) 
    \label{eq:calForce}
\end{align}
where $E_j$ is the potential energy in the current (active) electronic state $j$, $m_a$ is the mass of nucleus $a$, and $\boldsymbol{R}_a$ is the Cartesian coordinates of nucleus $a$. The potential energies and the corresponding gradients (i.e., atomic forces) can be predicted using NequIP. On the other hand, the evolution of electronic degrees of freedom is determined by the time-dependent Schrödinger equation as
\begin{equation}  
    \dot{\rho}_{jk}=-\frac{i}{\hbar}\rho_{jk}(E_j-E_k)-\sum_{l}\boldsymbol{\dot{R}}\cdot(\boldsymbol{d}_{jl}\rho_{lk}-\rho_{jl}\boldsymbol{d}_{lk})
    \label{eq:evolution of electron density}
\end{equation}
where $\rho_{jk}$ is the element of electron density matrix related to state $j$ and $k$, and $\boldsymbol{d_{jk}}$ is the nonadiabatic coupling vector between state $j$ and $k$, which can be expressed as
\begin{equation}  
    \boldsymbol{d}_{jk}(\boldsymbol{R})=\left\langle\psi_j(\boldsymbol{R})\left|\boldsymbol{\nabla_R} \right|\psi_k(\boldsymbol{R}) \right\rangle
\end{equation}
The evolution of electron density matrix determines the transition of nuclear motion from state $j$ to $k$ according to a probability $P_{j\rightarrow k}$ as
\begin{equation}  
    P_{j\rightarrow k}(t,t+\Delta t)=\max\left[0,\frac{\int_{t}^{t+\Delta t} 2Re{(\boldsymbol{\dot{R}}\cdot\boldsymbol{d_{jk}}\rho_{kj} )}d\tau}{\rho_{jj}(t)} \right]
    \label{eq:hopping judge}
\end{equation}
For two-state systems simulated in this work, the transition possibility can be written as
\begin{equation}  
    P_{j\rightarrow k}(t,t+\Delta t)=\max\left[0,\frac{\rho_{jj}(t+\Delta t)-\rho_{jj}(t)}{\rho_{jj}(t)} \right]
    \label{eq:hopping judge two state}
\end{equation}

\subsection{LSTM model}
\begin{figure}[ht]
    \centering
    \begin{subfigure}[b]{0.30\textwidth}
        \centering
        \includegraphics[width=\textwidth]{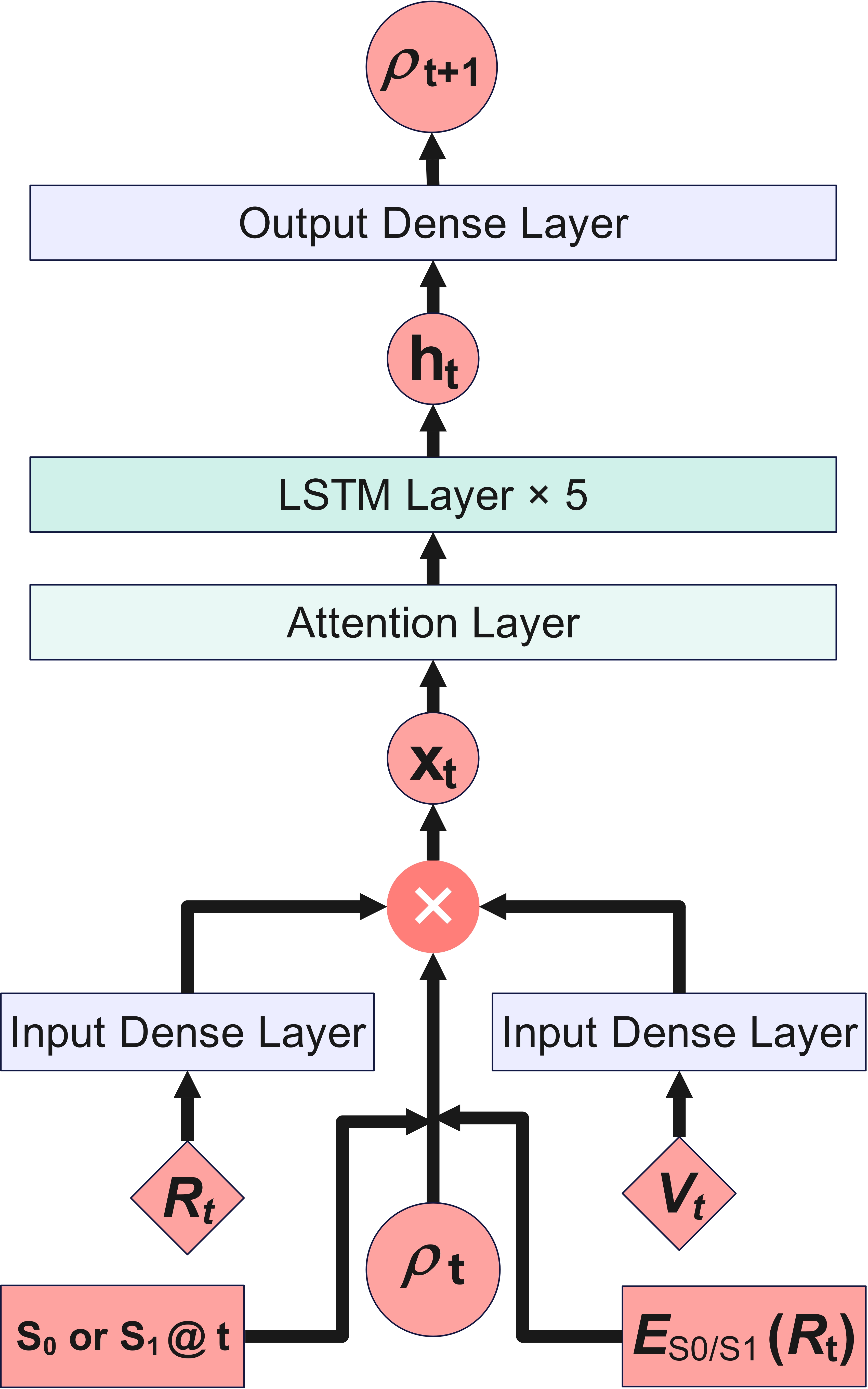}  
        \subcaption{}
        \label{fig:LSTMnetwork}
    \end{subfigure}
    \hfill
    \begin{subfigure}[b]{0.48\textwidth}
        \centering
        \includegraphics[width=\textwidth]{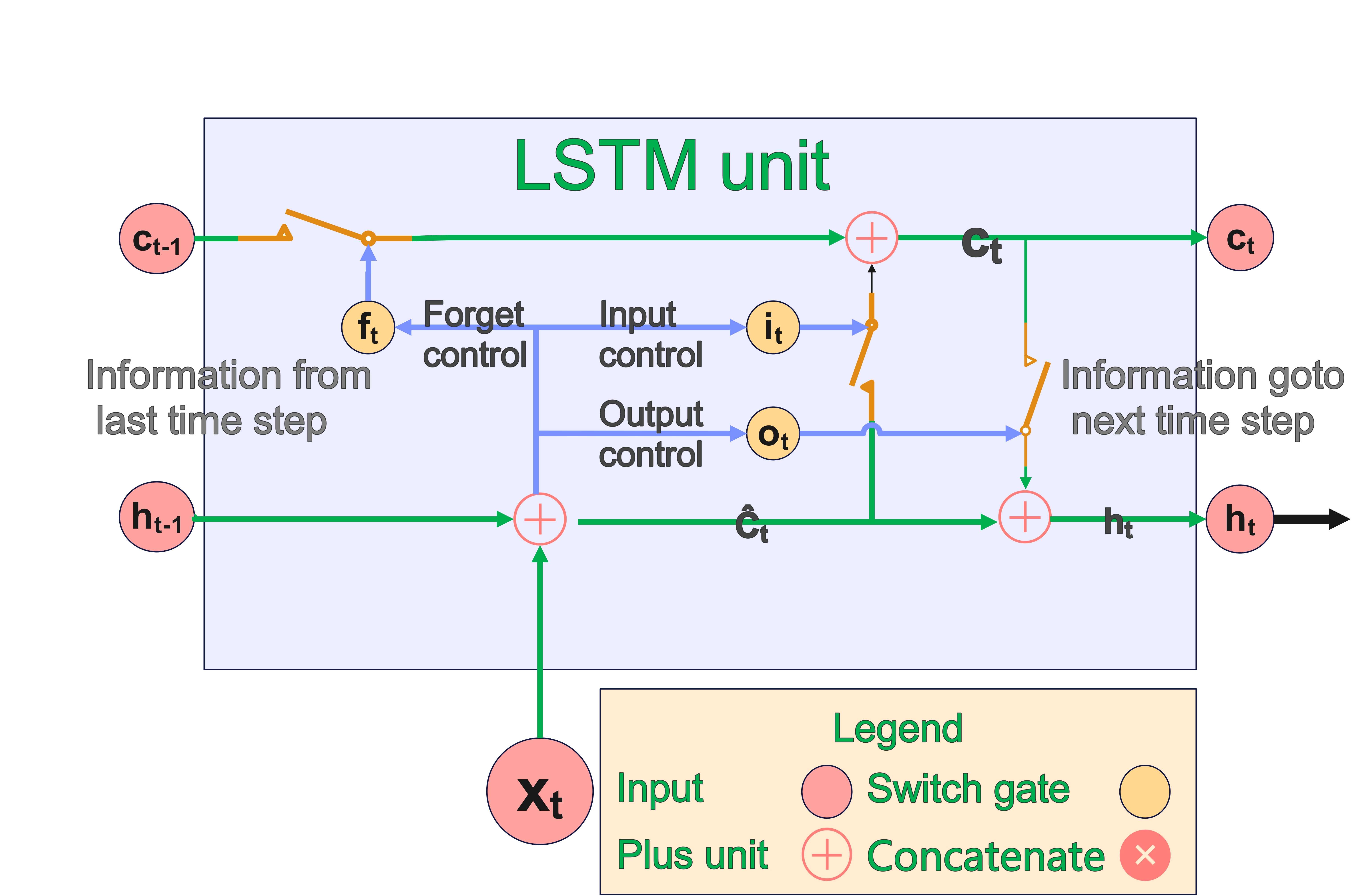}  
        \subcaption{}
        \label{fig:LSTMunit}
    \end{subfigure}
    \caption{Framework (a) and unit (b) of LSTM network model.}
    \label{fig:LSTMallmodel}
\end{figure}

LSTM can be considered as a functional formed by the composition of a specific mathematical operations. The total framework and a typical unit of the LSTM network model were shown in Figure \ref{fig:LSTMallmodel}. Based on the inputs $\boldsymbol{x}_t$ (processed via the first attention layer) in the current time step and the hidden variables $\boldsymbol{h}_{t-1}$ (or short-term state) in the last step, three gates that control the “memory” during time evolution are employed as
\begin{equation}  
    \boldsymbol{{f}_t} = \mathrm{sigmod}(\boldsymbol{W_{fx}x_t}+\boldsymbol{W_{fh}h_{t-1}}+\boldsymbol{B_f})
    \label{eq:gate}
\end{equation}
\begin{equation}  
    \boldsymbol{{i}_t} = \mathrm{sigmod}(\boldsymbol{W_{ix}x_t}+\boldsymbol{W_{ih}h_{t-1}}+\boldsymbol{B_i})
\end{equation}
\begin{equation}  
    \boldsymbol{{o}_t} = \mathrm{sigmod}(\boldsymbol{W_{ox}x_t}+\boldsymbol{W_{oh}h_{t-1}}+\boldsymbol{B_o})
\end{equation}
{
Where forget gate $\mathbf{f}_{t}$ is designed to decide keeping previous memory, input gate $\mathbf{i}_{t}$ decides inclusion of current information, and output gate $\mathbf{o}_{t}$ updates the hidden variables.
Then the hidden variables $\boldsymbol{h_t}$ and memory cells $\boldsymbol{c_t}$ (or long-term state) in the current step are calculated as 
\begin{equation}  
    \boldsymbol{\widetilde{c}_t} = \mathrm{tanh}(\boldsymbol{W_{cx}x_t}+\boldsymbol{W_{ch}h_{t-1}}+\boldsymbol{B_c})
\end{equation}
\begin{equation}  
    \boldsymbol{c_t} = \boldsymbol{{f}_t} \odot \boldsymbol{c_{t-1}}  + \boldsymbol{i_t} \odot \boldsymbol{\widetilde{c}_t}
\end{equation}
\begin{equation}  
    \boldsymbol{h_t} =  \mathrm{tanh}(\boldsymbol{c_t}) \odot \boldsymbol{\widetilde{o}_t}
    \label{eq:netoutput}
\end{equation}
where $\odot$ denotes pointwise multiplication. 
All weight parameters $\mathbf{W}$ and bias parameters $\mathbf{B}$ are freely optimized at training stage. 
Their dimensionality depends on input and output vectors.
In this work, five LSTM layers are employed as follows:
\begin{equation}  
    \left\{ \mathbf{h}_{t}^{[l]},\mathbf{c}_{t}^{[l]}\right\} = LSTM\left(\mathbf{h}_{t}^{[l-1]},\mathbf{h}_{t-1}^{[l]},\mathbf{c}_{t-1}^{[l]};\mathbf{W^{\mathrm{[\mathit{l}]}}},\mathbf{\mathbf{B^{\mathrm{[\mathit{l}]}}}}\right)
\end{equation}
when $l>1$.
Finally, the electron density matrix $\boldsymbol{\rho_t}$ is obtained from $\boldsymbol{h_t}$ via dense layers, which are denoted as “Output Dense Layer” in Figure 1, as follows
\begin{equation}  
    \boldsymbol{\rho_{t+1}} = f(\boldsymbol{W_{yh}^{[l]}h_t^{[l]}}+\boldsymbol{B_y^{[l]}})
    \label{eq:youtput}
\end{equation}
}
where $f$ is an active function such as the sigmoid, tanh and rectified linear unit (ReLu) functions. In this work, we also built encoders in prior to the LSTM layers for dimensionality reduction extraction of input features. They are denoted as “Input Dense Layer” for simplicity in Figure \ref{fig:LSTMallmodel}. The detailed structures of LSTM framework and hyperparameters used in this work can be seen in Figure S1 and Table S1.

\subsection{Input Features}
How to represent the nuclear and electronic degrees of freedom with appropriate input features is the key to enhance the performance of LSTM predictions. For a low-dimensional system such as Tully's three models, it is usually sufficient to apply some fundamental input features to LSTM, such as electron densities, potential energies, nuclear coordinates and velocities. Modeling on realistic molecules is much more difficult. Here we proposed five groups of input features to simulate realistic photochemical reactions: atomic coordinates, atomic velocities, elements of electron density matrix, potential energy difference between electronic states and index of active electronic state.

\subsubsection{SOAP Descriptors}
The information of nuclear degrees of freedom is represented in the first and second groups of input features. The smooth overlap of atomic orbitals (SOAP) descriptor\cite{SOAP.PhysRevB.87.184115,HIMANEN2020106949}, which provides an elegant way for translation, rotation and permutation invariance of atoms, was employed in this work.
{
The basic idea of the SOAP descriptor is to define the atomic density of the space with Gaussian functions.
The density of an element with nuclear charge $Z$ is
\begin{equation}  
    \rho^Z(\boldsymbol{r})=\sum_{i}^{N_Z}e^{-\frac{1}{2\sigma^2}|\boldsymbol{r-R_i}|^2}
    \label{eq:SOAPchargedefine}
\end{equation}
where $N_Z$ is number of atoms with atomic charge $Z$, $\boldsymbol{R_i}$ is the coordinate of atom $i$ with nuclear charge $Z$, and $\sigma$ is a pre-defined hyperparameter.
Details of hyperparameter used in this work are listed in the Supplemental Information. 
The density can be expanded with with a set of spherical harmonic ($Y_{lm}$) and radial Gaussian ($g_n$) basis functions:
\begin{equation}  
    \rho^Z(\boldsymbol{r})=\sum_{nlm}c_{nlm}^{Z}g_n(r)Y_{lm}(\theta,\varphi)
\end{equation} 
where $c_{nlm}^Z$ is the solved coefficient, which is translation, rotation and permutation invariant. The final SOAP descriptors are constructed as
\begin{equation}  
    p_{nn'l}^{Z^1,Z^2} = \pi \sqrt{\frac{8}{2l+1}}\sum_{m}(c_{nlm}^{Z^1})^*(c_{n'lm}^{Z^2})
\end{equation}}
In this work, we built a set of SOAP descriptors directly to represent nuclear coordinates. The number of SOAP descriptors was set as $N \times N_{\text{SOAP}}$, where $N$ is the total number of atoms. We also changed $\boldsymbol{r}$ and $\boldsymbol{R}_i$ in eq \ref{eq:SOAPchargedefine} to $\boldsymbol{v}$ and $\boldsymbol{V}_i$, respectively, and built another set of input features to describe nuclear velocities in each dynamic step. 
{
It should be noted that our choice to encode velocities using the SOAP descriptor was not physically-inspired. Instead, it was a mathematical strategy to ensure the translation and rotation invariance of the velocity input features within the LSTM framework.
}
The min-max normalization was performed, in which the maximum and minimum values of each feature were derived from the training set. The normalized values were subsequently applied to the input dense layers for dimensionality reduction. In this work, we decreased the number of input features in the first group of as well as that in the second group from $N \times N_{\text{SOAP}}$ to a smaller $N_{\text{nuc}}$.

\subsubsection{Electronic State Density}
\begin{figure*}[t]
    \centering
    \begin{minipage}{0.9\textwidth}
        \centering
        \begin{equation}
            g(x) = 
            \begin{cases} 
            -\dfrac{0.05}{0.55 + (x-0.5)}, & \text{if } x \leq 0.1 \\[2ex]
            \dfrac{19375}{576} (x-0.5)^5 - \dfrac{925}{144} (x-0.5)^3 + (x-0.5), & \text{if }  0.1 \leq x \leq 0.9 \\[2ex]
            \dfrac{0.05}{0.55 - (x-0.5)}, & \text{if } x \geq 0.9
            \end{cases}
            \label{eq:g_xx}
        \end{equation}
    \end{minipage}
    \label{eq:g_x}
\end{figure*}

The information of electronic degrees of freedom is represented explicitly with electron density matrix. As discussed in our previous work, the diagonal elements of density matrix, which are relevant to the population of the simulated system on different electronic states, play the most essential role on the prediction performance of LSTM.
{
On one hand, success of LSTM-driven trajectory is heavily dependent on correct attention to the starting point of coherence in density. On the other hand, the activation functions in neural networks are usually activated around $0$. Direct usage of diagonal elements of density matrix as input is unrecommended. Instead, we introduced a rescaling function $g$ as eq \ref{eq:g_xx}, where $x$ denotes a diagonal element of density matrix. 
Detailed explanation of eq \ref{eq:g_xx} is shown in Supporting Information.
}
No rescale is applied to off-diagonal elements of density matrix since their characteristics is not difficult for training, at least in the present cases.

\subsubsection{Potential Energy}
The fourth and fifth groups of input features include more information of simulated systems during time propagation in an implicit way. The potential energy difference has a great influence on the evolution during surface hopping dynamics, especially once the transition event takes place. We applied $x_{log\Delta E}$ and $x_{ {\Delta E}^{-1}}$ as the fourth group of input features to capture the difference between $E_{\mathrm{S}_1}$ and $E_{\mathrm{S}_0}$ as
\begin{equation}  
    \begin{cases}   
    x_{log\Delta E}=\log_{10}( E_{\mathrm{S}_1}-E_{\mathrm{S}_0} + \epsilon )-1  \\
    x_{ {\Delta E}^{-1}}=\frac{2.5}{ ( E_{\mathrm{S}_1}-E_{\mathrm{S}_0}) +1 } -1 
    \end{cases}  
    \label{E remap}
\end{equation}
where $E_{\mathrm{S}_1}$ and $E_{\mathrm{S}_0}$ denote the potential energies in the $\mathrm{S}_1$ and $\mathrm{S}_0$ state, respectively, in unit of kcal/mol. $\epsilon$ is a small positive constant introduced to prevent numerical instability. The active electronic state is another key to surface hopping. In this work we simulated two-state systems, and thus the index of active state was set to -1 for the ground state and 1 for the excited state as the fifth group of input features.

\subsubsection{Physical Constraint}
In brief, the time-dependent information of nuclear and electronic degrees of freedom in several previous steps is embedded into input features of the LSTM model, and the elements of electron density matrix in the current step is predicted by LSTM. For two-state systems simulated in this work, the outputs of LSTM include $\rho_{00}$, $\rho_{11}$, Re($\rho_{01}$) and Im($\rho_{01}$). Until now, the physical constraints to electron density matrix are not considered during the modeling of LSTM. We addressed this issue after output dense layers of LSTM. First, the value of diagonal element should be recapped once it exceeds the range between 0 and 1, that is
\begin{equation}  
    \rho_{jj}  \text{:=}   
    \begin{cases}   
        0 & \text{if } \rho_{jj} \leq 0\\
        1 & \text{if } \rho_{jj} \geq 1
    \end{cases}  
    \label{eq:RhoconstraintA}
\end{equation}

Second, the trace of density matrix should be a constant of 1, leading to normalization to the predicted values as
\begin{equation}  
    \rho_{jj}  \text{:=}  \frac{\rho_{jj}}{ \sum_k \rho_{kk}  }
    \label{eq:RhoconstraintB}
\end{equation}
Finally, the module of off-diagonal elements should match the module of diagonal elements. We have observed that violation of this constraint may result in some errors in the judgment of surface hopping during long-term propagation. The following correction for a two-state system is performed as
\begin{equation}  
    \begin{cases}   
    Re(\rho_{01}) \text{:=} Re(\rho_{01})\sqrt{\frac{\rho_{00}\rho_{11}}{|M_{01}|^2}  } \\
    Im(\rho_{01}) \text{:=} Im(\rho_{01})\sqrt{\frac{\rho_{00}\rho_{11}}{|M_{01}|^2}  } \\
    \end{cases}
    \label{eq:RhoconstraintC}
\end{equation}
where $\rho_{00}$ and $\rho_{11}$ have been corrected in eqs \ref{eq:RhoconstraintA} and \ref{eq:RhoconstraintB}, and $|M_{01}|$ is the module of $\rho_{01}$ before correction.

\subsection{Work procedure}
\begin{figure}[ht] 
    \centering 
    \includegraphics[width=0.47\textwidth]{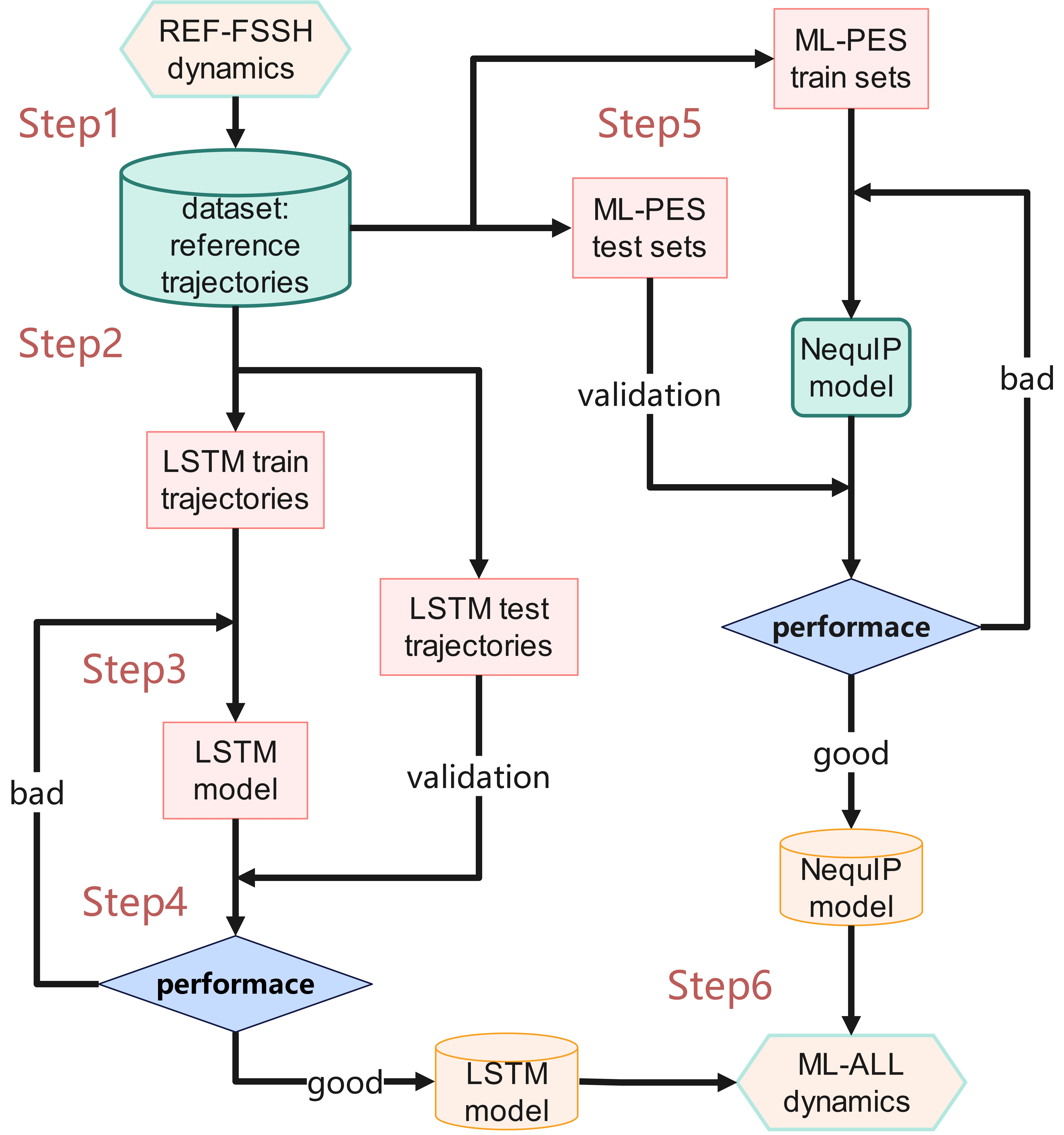}   
    \caption{Workflow of combined LSTM and NequIP for FSSH simulations.}   
    \label{fig:workflow} 
\end{figure}
The whole process of this work was illustrated in Figure \ref{fig:workflow} and summarized as follows: 

(1) Perform conventional FSSH simulations on the selected photochemical reaction system with different random seeds. A small number of reference trajectories were generated (e.g., 9 trajectories for $\mathrm{CH_2NH}$ and 10 trajectories for azobenzene) and further divided into training and test representative trajectories. For each trajectory, the atomic coordinates and velocities, electron density matrix, potential energies in all relevant electronic states as well as the gradients with respect to atomic coordinates, and the active electronic state in each dynamic step were recorded for ML modeling. 

(2) Extract a time series of input features with $M$ frames from one training trajectory as a sequence to build the LSTM model. Similar to our previous work, the sequence is shifted forward by $M_{\text{shift}}$ frames until the end of this trajectory. This step is repeated across other training trajectories to extract a large number of sequences (e.g., 8 trajectories for $\mathrm{CH_2NH}$ yields 16,000 sequences and 9 trajectories for azobenzene yields 13,500 sequences). 

(3) Construct an LSTM model based on the extracted sequences with a standard machine learning modeling procedure.

(4) Use the test trajectories (e.g., one representative trajectory for $\mathrm{CH_2NH}$) to validate LSTM. For each selected trajectory, the initial $M$-length sequence consisting of all input features is applied to the constructed LSTM model, predicting the electron density matrix in the next step. Then the next $M$-length sequence, in which only the electron density matrix is replaced with the prediction result, is applied to LSTM iteratively. The LSTM model is acceptable if the LSTM-driven time evolution of electron density matrix is consistent with the corresponding reference result. {Otherwise, go back to step 3 and rebuild LSTM using different hyperparameters.}

(5) Extract thousands of snapshots from all reference trajectories to build training and test sets of NequIP (e.g., 2,000 snapshots for $\mathrm{CH_2NH}$ from 9 trajectories). The snapshots populated in different electronic states are separated to build individual NequIP models for adiabatic PESs in each state. If the training loss or test MAE of the ML-PES model is unsatisfactory, we can consider expanding the snapshot dataset, adjusting the weights of energy and force terms in the loss function, or performing additional targeted molecular dynamics simulations to sample new configurations in underrepresented regions.

(6) Combine the constructed NequIP and LSTM models to perform FSSH simulations on the selected photochemical reaction system. Nuclear degrees of freedom are still propagated according to the Newton's second law, but the potential energies and atomic forces are predicted with NequIP, avoiding expensive excited-state electronic structure calculations. Time evolution of electronic degrees of freedom are driven by LSTM without the need of nonadiabatic couplings. Collective results such as excited-state lifetime and photochemical product yield can be estimated based on hundreds or thousands of simulated trajectories.

There are two main differences between our previous work and this study. First, we introduced NequIP models to construct adiabatic PESs in all relevant electronic states, which is critical to save expensive computational cost on electronic structure calculations of realistic molecular systems. It should be noted that the test trajectories selected for validating LSTM should exhibit representative changes on density matrix, while the dataset for building NequIP should cover broad regions on PESs. It means that more dynamics trajectories may be required in order to obtain more accurate ML-PESs, especially for large molecules. Second, the validation of LSTM in step 4 changes from “on-the-fly” to “off-line”. In our previous study on Tully's three models, LSTM-FSSH dynamic simulations were implemented under initial conditions and random numbers as the same as test trajectories. Time evolution of nuclear degrees of freedom would be affected by LSTM once a hopping event occurs. In this work, however, only the electron density matrix evolves with LSTM, while the atomic coordinates and velocities remain the same as those in the reference FSSH trajectories. Since the transition probability is determined by $\rho_{jj}$, the active state should be also influenced by LSTM, changing nuclear motion sequentially. Here we ignored such influence for simplicity as follows: when the index of active state in an LSTM-driven trajectory becomes different from its reference (i.e., representative trajectory), which is due to prediction errors on $\rho_{jj}$, the index of active state as the fifth group of input features would be determined by LSTM, but the nuclear motion (i.e., the first, second and fourth input features) would be subjected to the reference trajectory.

\subsection{Simulation details}
\begin{figure*}[ht]
    \centering
    \begin{subfigure}[b]{0.42\textwidth}
        \centering
        \includegraphics[width=\textwidth]{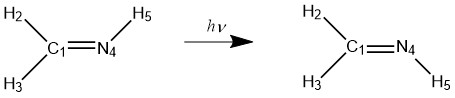}
        \caption{}
        \label{fig:ReactionMet}
    \end{subfigure}
    \hfill
    \begin{subfigure}[b]{0.50\textwidth}
        \centering
        \includegraphics[width=\textwidth]{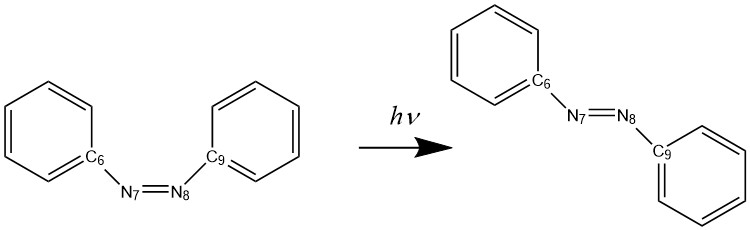}
        \caption{}
        \label{fig:ReactionAzo}
    \end{subfigure}
    \caption{Photoisomerizations of $\mathrm{CH_2NH}$ (a) and azobenzene (b).}
    \label{fig:Isomerization}
\end{figure*}

The first test system is the photoisomerization of $\mathrm{CH_2NH}$ along the central dihedral angle, starting from its $\mathrm{S}_1$ state. In this work, all initial molecular configurations were defined artificially as the $cis$-isomer, and the $cis$- and $trans$-isomers were distinguished by a hydrogen atom labeled as $\mathrm{H5}$ (see Figure \ref{fig:ReactionMet}). We employed the complete active space self-consistent field (CASSCF) method with the basis set of 6-31G(d) for electronic structure calculations to generate reference FSSH dynamic trajectories in step 1. The active space consists of 2 electrons and 2 orbitals ($\pi$ and $\pi^*$). The second test system is the $cis$-to-$trans$ photoisomerization of azobenzene along $\angle \mathrm{C}$-$\mathrm{N}$-$\mathrm{N}$-$\mathrm{C}$, starting from its $\mathrm{S}_1$ state (see Figure \ref{fig:ReactionAzo}). We applied OM2 Hamiltonian with the multi-reference configuration interaction (OM2/MRCI) method for electronic structure calculations to generate reference trajectories. An active space of 10 electrons in 8 orbitals was used, including four $\pi$ orbitals, one $n$ orbital and three $\pi^*$ orbitals. 
{
Although the computational overheads of OM2/MRCI and NequIP are similar for the test systems, our ML approach can be directly applied to more expensive electronic structure approaches. Note that the relationship between the computational level of electronic structure method and the accuracy of dynamics simulation results is not monotonic for realistic molecular systems.}
The OpenMolcas\cite{OpenMolcas2023} and MNDO programs\cite{Thiel2019MNDO} were used for CASSCF and OM2/MRCI\cite{Weber1996,Weber2000} calculations, respectively.

For each test system, 1,000 initial conditions were generated based on the Wigner sampling in 300 K using the corresponding module in the Newton-X program\cite{10.1021/acs.jctc.2c00804}. The integration time steps of the nuclear and electronic motions were set as 0.1 and $5\times 10^{-4}$ fs, respectively, for all nonadiabatic dynamic simulations. Total simulation time of each trajectory was 400 and 300 fs for $\mathrm{CH_2NH}$ and azobenzene, respectively. No decoherence correction was applied to FSSH in this work for simplicity. LSTM models were trained using Keras v2.13.1\cite{Chollet2015Keras} combined with TensorFlow v2.13.0\cite{Abadi2015TensorFlow} . Adiabatic PESs of these two systems were constructed with NequIP v0.6.1\cite{Batzner2022,Geiger2021}.

\section{RESULTS AND DISCUSSION}
\label{sec:RESULTSANDDISC}
\subsection{Performance of NequIP models}
\begin{figure*}[ht]
    \centering
    \begin{subfigure}[b]{1.6in}
        \centering
        \includegraphics[width=\textwidth]{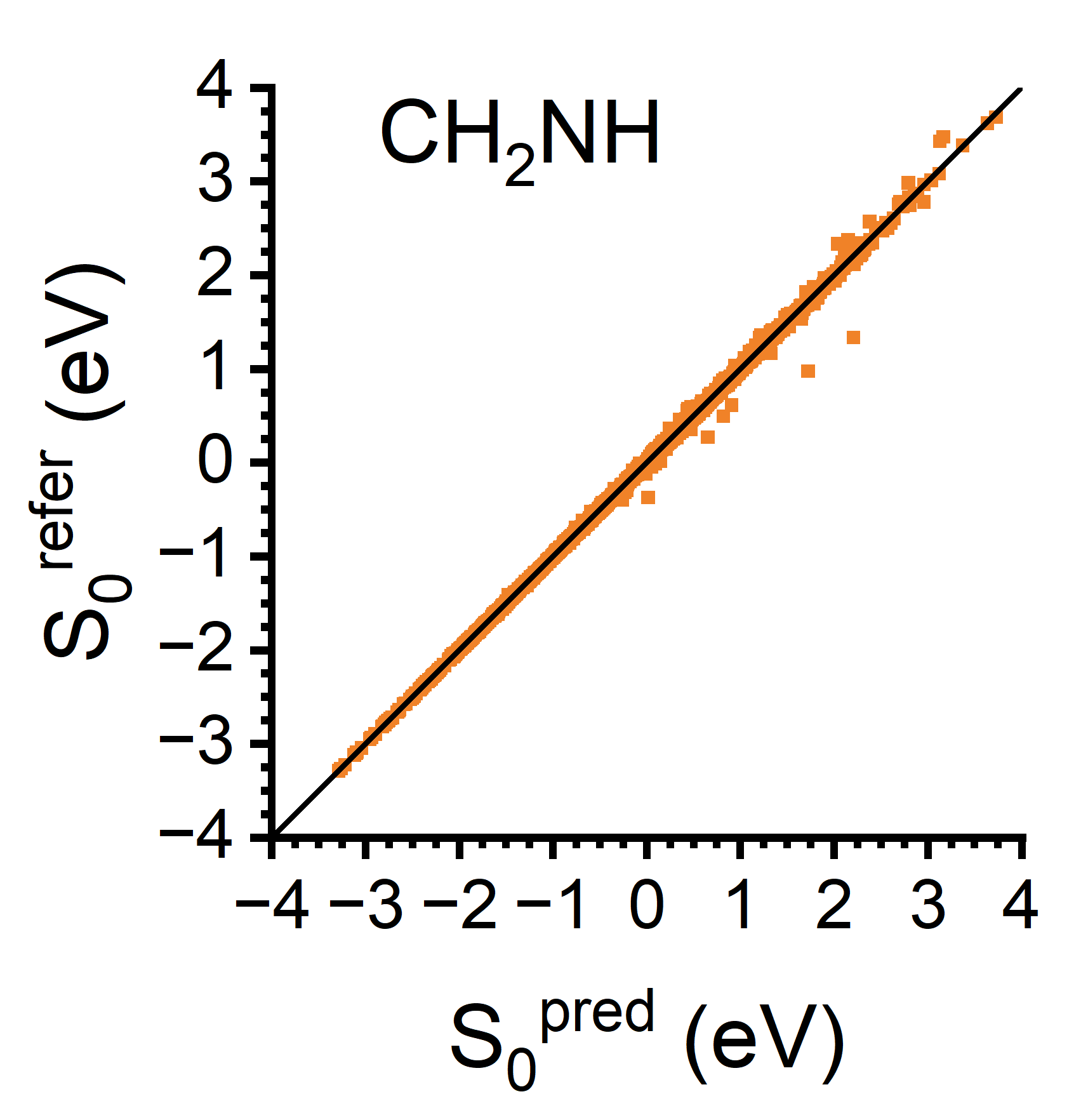}
        \caption{}
        \label{fig:MetNequIPA}
    \end{subfigure}
    \hfill
    \begin{subfigure}[b]{1.6in}
        \centering
        \includegraphics[width=\textwidth]{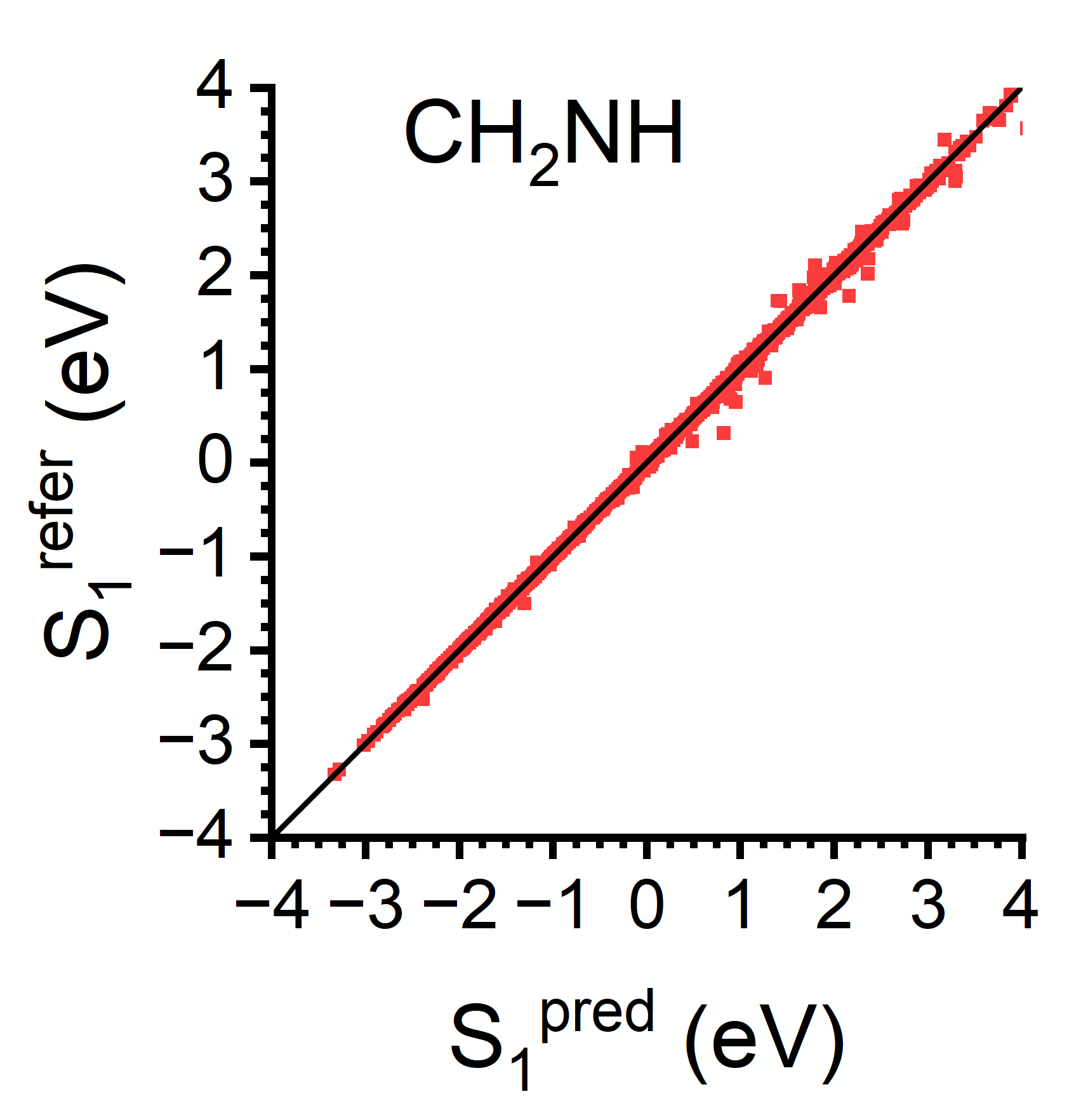}
        \caption{}
        \label{fig:MetNequIPB}
    \end{subfigure}
    \hfill
    \begin{subfigure}[b]{1.6in}
        \centering
        \includegraphics[width=\textwidth]{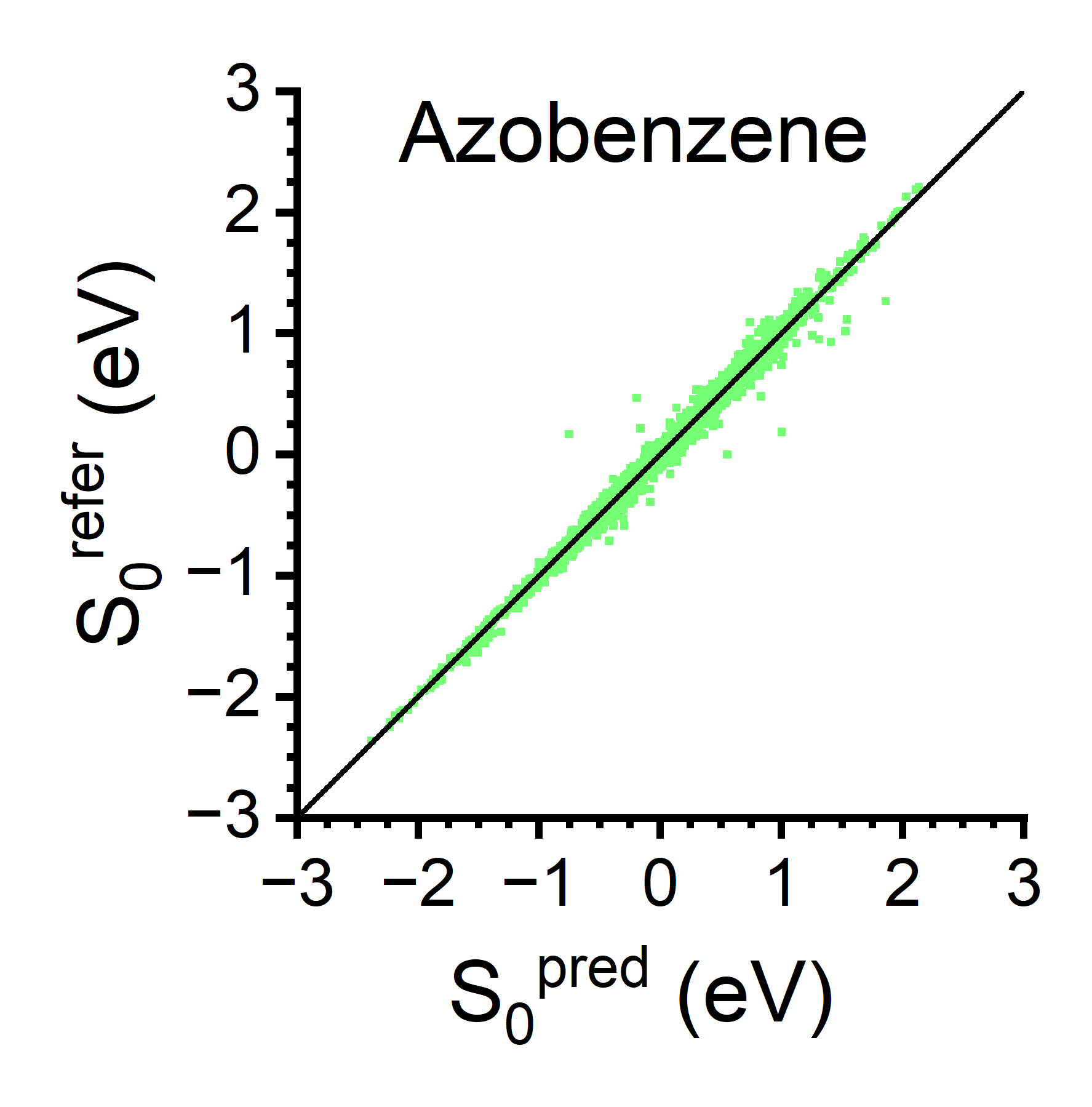}
        \caption{}
        \label{fig:AzoNequIPA}
    \end{subfigure}
    \hfill
    \begin{subfigure}[b]{1.6in}
        \centering
        \includegraphics[width=\textwidth]{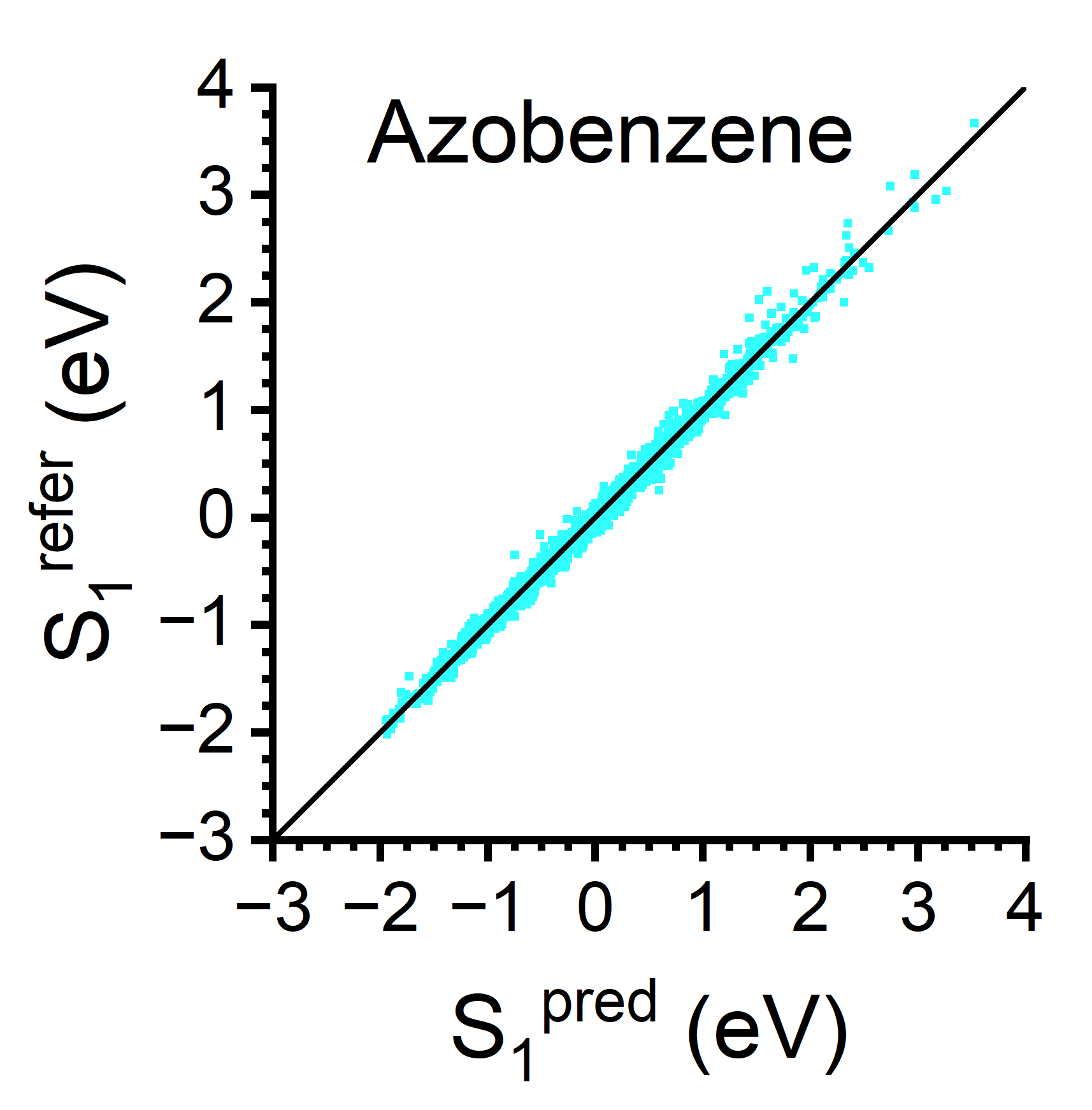}
        \caption{}
        \label{fig:AzoNequIPB}
    \end{subfigure}
    \caption{Comparisons between reference and NequIP-predicted potential energies for $\mathrm{CH_2NH}$ in $\mathrm{S}_0$ (a), $\mathrm{CH_2NH}$ in $\mathrm{S}_1$ (b), azobenzene in $\mathrm{S}_0$ (c), and azobenzene in $\mathrm{S}_1$ (d). Absolute values of potential energies have been shifted by a constant for clarity: 2,555 eV for $\mathrm{CH_2NH}$ in $\mathrm{S}_0$, 2,552 eV for $\mathrm{CH_2NH}$ in $\mathrm{S}_1$, 2,102 eV for azobenzene in $\mathrm{S}_0$, and 2,100 eV for azobenzene in $\mathrm{S}_1$.}
    \label{fig:AllNequIP}
\end{figure*}

We extracted 4,000 snapshots from the reference trajectories of $\mathrm{CH_2NH}$ and split them by 1:1 to generate the training and test sets to build the potential energy surfaces in its $\mathrm{S}_0$ and $\mathrm{S}_1$ states. {
Comparison between reference and predicted potential energies are shown in Figures \ref{fig:MetNequIPA} and \ref{fig:MetNequIPB}.
The test mean absolute errors (MAEs) of $E(\mathrm{S}_0)$ and $E(\mathrm{S}_1)$ were 0.016 and 0.019 eV, respectively, with the force loss in the order of $10^{-2}$ eV/\AA.
The prediction accuracy can be accepted since it's well within the threshold of chemical accuracy as 1 kcal/mol ($\sim $0.043 eV).
} Key configurations observed in the reference trajectories, involving $cis$- and $trans$-isomers as well as minimum energy conical intersections, were shown in Figure S5a. Notably, a nitrene in which the H5 atom moves from nitrogen to carbon (see Figure S5b) also appeared in a few cases. It is more difficult for ML to accurately predict potential energies of such nitrene-like configurations in the high-energy region of PES. Although the performance of NequIP was robust, the influence of these nitrene-like configurations on dynamics simulations cannot be ignored at all times. We will discuss this issue later.

Because of the high-dimensional complexity of azobenzene, building accurate deep learning PESs is more challenging. To ensure sufficient sampling of the configurational space, we performed 20 more FSSH simulations at the OM2/MRCI computational level and extracted 4,000 snapshots from a total 30 reference trajectories. 
{
These trajectories include both successful isomerizing events and non-isomerizing paths, with a configuration ratio of approximately 63:37 for \textit{cis} and \textit{trans} isomers.}
These snapshots were split to training and test sets to construct PESs. Key configurations related to photoisomerization of azobenzene can be seen in Figure S5c.
{
Although the crossing topology between ground state and excited state surfaces are qualitatively reproduced, it should be noted that potential energy of the \textit{cis} isomer is underestimated, which is a known disadvantage of the semiempirical methods such as OM2 and AM1.\cite{Thiel11_1506}}
Comparison between reference and predicted potential energies are shown in Figure \ref{fig:AzoNequIPA} and \ref{fig:AzoNequIPB}. The MAEs of $E(\mathrm{S}_0)$ and $E(\mathrm{S}_1)$ on the test sets were 0.052 and 0.055 eV, respectively. The corresponding force loss per atom was increased to the order of $10^{-1}$ eV/\AA. 
Although the MAEs are slightly higher than those for $\mathrm{CH_2NH}$, they remain near the chemical accuracy threshold.
On one hand, a larger dataset required to construct accurate PESs of azobenzene retards the efficiency of the whole procedure. With the development of deep learning potentials, however, we believe that the need of big data can be reduced using the pre-train-and-fine-tune strategy in the near future. On the other hand, 10 reference trajectories were sufficient for LSTM training on azobenzene (see below), which indicates potential superiority of LSTM on large molecular systems. 

\subsection{Representative trajectories}
\label{subsec:Representative trajectory test}
\begin{figure*}[ht]
    \centering
    \begin{subfigure}[b]{0.45\textwidth}
        \centering
        \includegraphics[width=\textwidth]{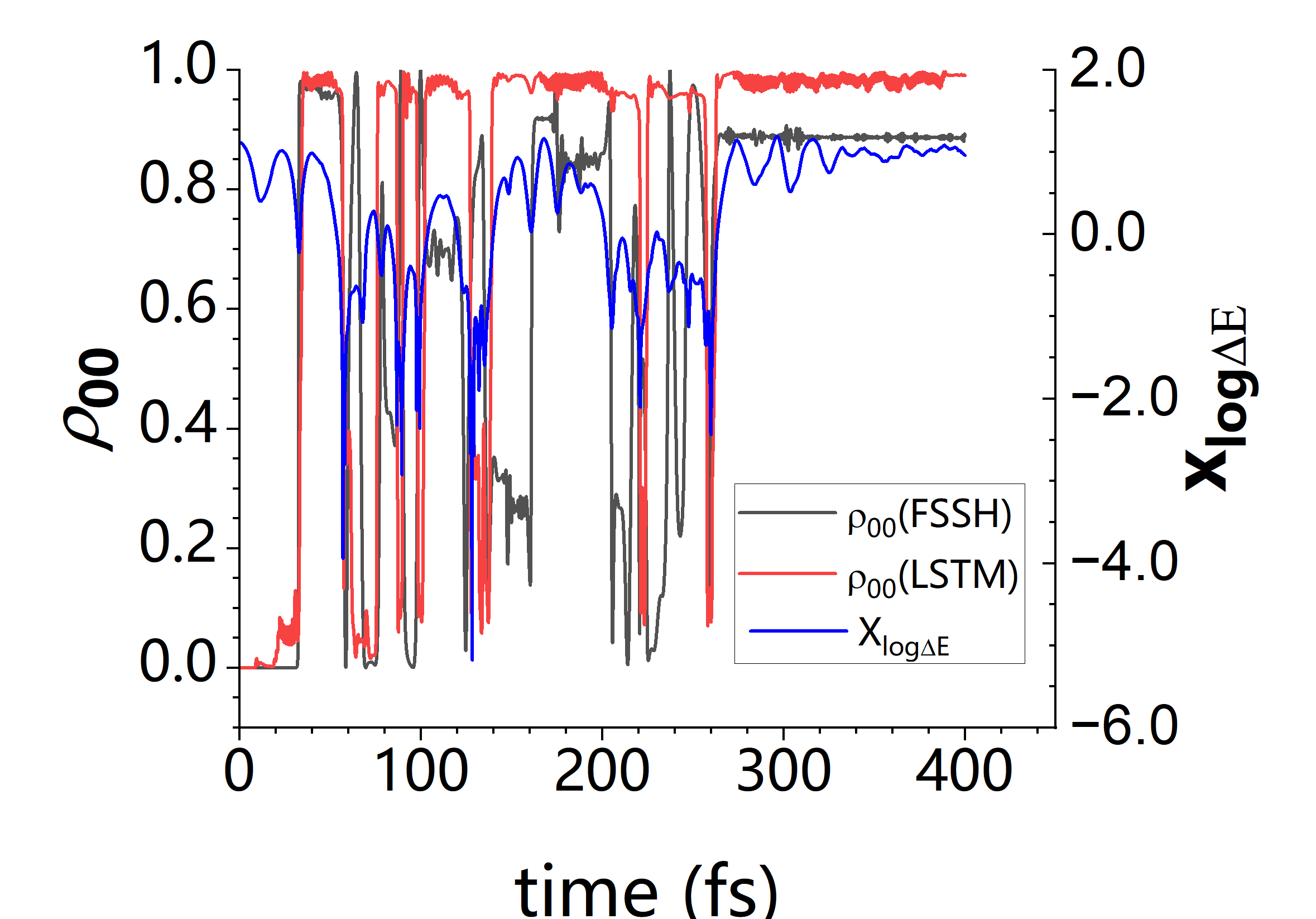}
        \caption{}
    \end{subfigure}
    \hfill
    \begin{subfigure}[b]{0.45\textwidth}
        \centering
        \includegraphics[width=\textwidth]{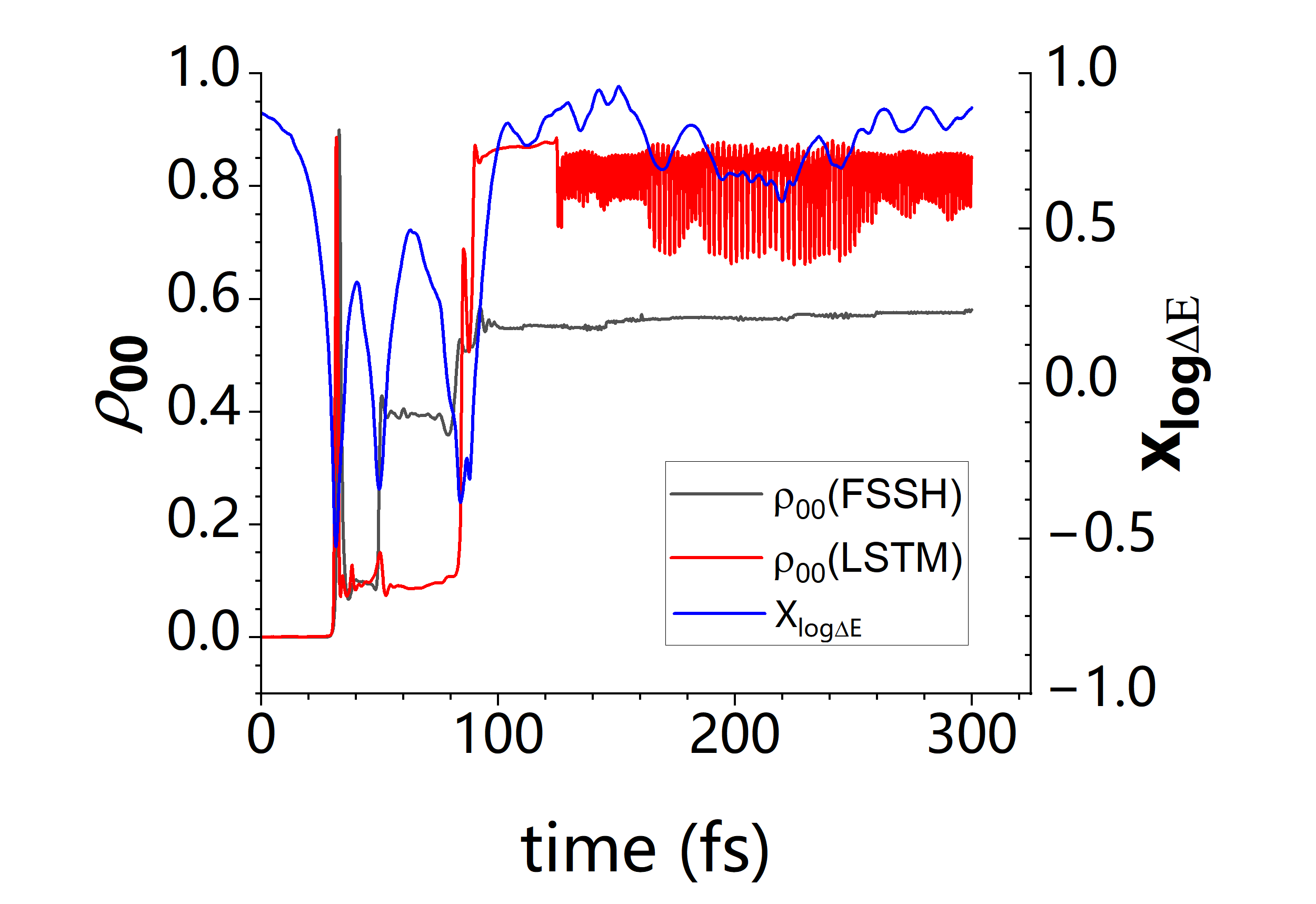}
        \caption{}
    \end{subfigure}
    \caption{Population of $\rho_{00}$ and $x_{log\Delta E}$ as functions of time evolved in representative trajectories for $\mathrm{CH_2NH}$ (a) and azobenzene (b). Different colors represent different variables: $\rho_{00}$ (FSSH) in black, $\rho_{00}$ (LSTM) in red, and $x_{log\Delta E}$ in blue.}
    \label{fig:Allrepresentative}
\end{figure*}
It has been observed previously that comparison between reference and predicted sequences is insufficient to validate the performance of LSTM because of the accumulated prediction errors during time evolution. Here we selected representative test trajectories for two test systems and examined time evolution of $\rho_{00}$ driven by conventional FSSH or constructed LSTM. The time-dependent values of $x_{log\Delta E}$ during LSTM-driven simulations were also detected because this input feature is relevant to a threshold of surface hopping. Any transition event would be rejected if the energy gap was larger than 10 kcal/mol, corresponding to a positive value of $x_{log\Delta E}$. 
{
While this threshold is suitable for the current systems, we acknowledge its potential limitations for reactions involving possible significant decay at larger energy gaps, for which the LSTM model should be retrained using a supplemental reference trajectory.} In contrast, a very negative $x_{log\Delta E}$ usually suggests strong coupling between two electronic states.

The simulation results of $\mathrm{CH_2NH}$ can be seen in Figure \ref{fig:Allrepresentative}. We mainly focused on the significant coupling within the time intervals of 30-36, 52-60 and 75-105 fs. First of all, consistency between reference and LSTM trajectories around the time period of the first surface hopping is essential. As shown in Figure \ref{fig:Allrepresentative}, both trajectories start in the $\mathrm{S}_1$ state and firstly switch to the ground state at about 35 fs, exhibiting the same pattern of time evolution of $\rho_{00}$. Then the reference trajectory undergoes three frequent transitions between 52 and 60 fs, finally returning to $\mathrm{S}_1$. LSTM captures the second but misses the third and fourth hopping events. Within the time interval of 75-105 fs, the values of $\rho_{00}$ strongly oscillate in both simulations, accompanied by an increasing deviation of LSTM from the reference data. Nevertheless, electron density matrix obtained using LSTM is comparable to the reference in the early stage of dynamic simulation. The overall performance of LSTM after 105 fs is still acceptable. Notably, the overestimation of $\rho_{00}$ between 130 and 140 fs may result in some decrease in simulated excited-state lifetime. Both trajectories converge to $\mathrm{S}_0$ at the end of simulations.

Photoisomerization of azobenzene is a more challenging system for LSTM training because of its higher dimensionality. Similar to $\mathrm{CH_2NH}$, more attention should be still paid to electronic motion in presence of strong coupling, especially around transition events in the early stage of simulations. As shown in Figure \ref{fig:Allrepresentative}, the reference trajectory approaches the intersection vicinity at about 30, 50 and 85 fs, leading to a sudden change of $\rho_{00}$ as well as surface hopping. LSTM reproduces the first two transitions occurring between 28 and 36 fs. Both trajectories switch from $\mathrm{S}_1$ to $\mathrm{S}_0$ and then back to $\mathrm{S}_1$. However, LSTM trajectory remain in the excited state within 47-53 fs, which deviates from the reference trajectory and may attribute to the overestimated excited-state lifetime. LSTM also correctly responses to the strong coupling around 80-90 fs. Despite the difference of converged $\rho_{00}$ between two simulations, both reference and LSTM trajectories decay to the ground state after 100 fs. 
The oscillations in $\rho_{00}$ observed between 150 and 300 fs is possiblity due to accumulated prediction errors, in long time simulations, as has been oberseved in our previous work\cite{doi:10.1021/acs.jpclett.2c02299}. 
{
Although these oscillations primarily manifest in physically non-sensitive regions with large energy gaps, they may potentially impact the reliability of dynamics on longer timescales for specific systems.} Training an LSTM model that remains perfectly stable over hundred femtosecond trajectory is exceptionally challenging, and future development on methods for decoherence may be necessary.
Nevertheless, the above performance of LSTM is acceptable for such a complicated system.

It should be noted that only a small portion of LSTMs are able to reproduce the reference data qualitatively in our practice. A majority of constructed models clearly fail to pass the above validation on representative test trajectories. Some negative examples were summarized in Figures S2 and S3.

\subsection{Collective results}
\begin{table*}[ht]
    \centering
    \begin{tabular}{l l c c c}
        \toprule
           & \textbf{System} & \textbf{REF-FSSH} & \textbf{ML-PES} & \textbf{ML-ALL} \\
        \midrule

        \addlinespace[1ex]
        \multirow{2}{*}{\textbf{product yield (\%)}} 
            & $\mathrm{CH_2NH}$    & 54.8 $\pm$ 1.1 & 49.9 $\pm$ 1.2 & 50.4 $\pm$ 1.2\\
            & Azobenzene          & 45.9 $\pm$ 0.8 & 43.1 $\pm$ 0.9 & 45.6 $\pm$ 0.9\\
        \addlinespace[1ex]

        \multirow{2}{*}{ $\mathbf{ \mathrm{S}_1 }$ \textbf{ lifetime (fs)}} 
            & $\mathrm{CH_2NH}$    & 70.4 $\pm$ 1.4 & 80.4 $\pm$ 1.9 & 68.8 $\pm$ 1.0 \\
            & Azobenzene          & 88.2 $\pm$ 1.3 & 84.2 $\pm$ 1.3 & 91.4 $\pm$ 1.3 \\
        \bottomrule
    \end{tabular}
    \caption{Predicted excited-state lifetimes and photochemical product yields of photoisomerizations of $\mathrm{CH_2NH}$ and azobenzene using different simulation methods.}
    \label{tab:Collectiveresults}
\end{table*}

We simulated independent dynamics trajectories under different initial conditions, i.e., 800 for $\mathrm{CH_2NH}$ and 1,000 for azobenzene, to obtain collective results. Three simulation methods denoted as REF-FSSH, ML-PES and ML-ALL were implemented individually. For each method, the total number of simulated trajectories has been enough to achieve converged collective results. As displayed in Figure S4, the excited-state lifetime collected from a smaller ensemble of 100 trajectories remains divergent. In comparison, our proposed LSTM requires only 10 reference trajectories for training but is able to produce hundreds of independent trajectories at very small cost. 

REF-FSSH provides reference data in absence of any ML model. ML-PES employs our constructed NequIP models to predict potential energies and gradients in the ground and excited states. The nonadiabatic coupling vector was still calculated using the external QM program. In other words, electronic motion was driven in the conventional way as REF-FSSH. ML-ALL is our proposed model that applies NequIP and LSTM to evolve nuclear and electronic degrees of freedom, respectively. Similar to our pervious observations, 6$\sim$8 \% of REF-FSSH trajectories failed because of unphysical dissociation of molecule. This problem appeared in 19\% of $\mathrm{CH_2NH}$ and 17\% of azobenzene trajectories when electronic structure calculation was replaced with NequIP. The former mainly originates from nitrene-like configurations, while the latter is caused by the difficulty in building ML-based PESs for large molecules. 
A detailed analysis about failed configurations for azobenzene is shown in Figure S6.
The failure percentage is still acceptable on account of much higher efficiency of ML predictions. {The event that violates the physical constraint on electronic density as $\rho_{00} + \rho_{11} = 1.0$ is rare (see Figure S7 and Table S2).} 

The excited-state lifetime and photochemical product yield were collected from successful trajectories. The $\mathrm{S}_1$ lifetime $\tau$ was obtained by fitting the time-dependent $\mathrm{S}_1$ population as $P(t)=H(t_0-t)+H(t-t_0 ) e^{-((t-t_0 ))/ \tau_d }$, resulting in $\tau=t_0+\tau_d$.

\begin{figure*}[ht]
    \centering
    \begin{subfigure}[b]{0.47\textwidth}
        \centering
        \includegraphics[width=\textwidth]{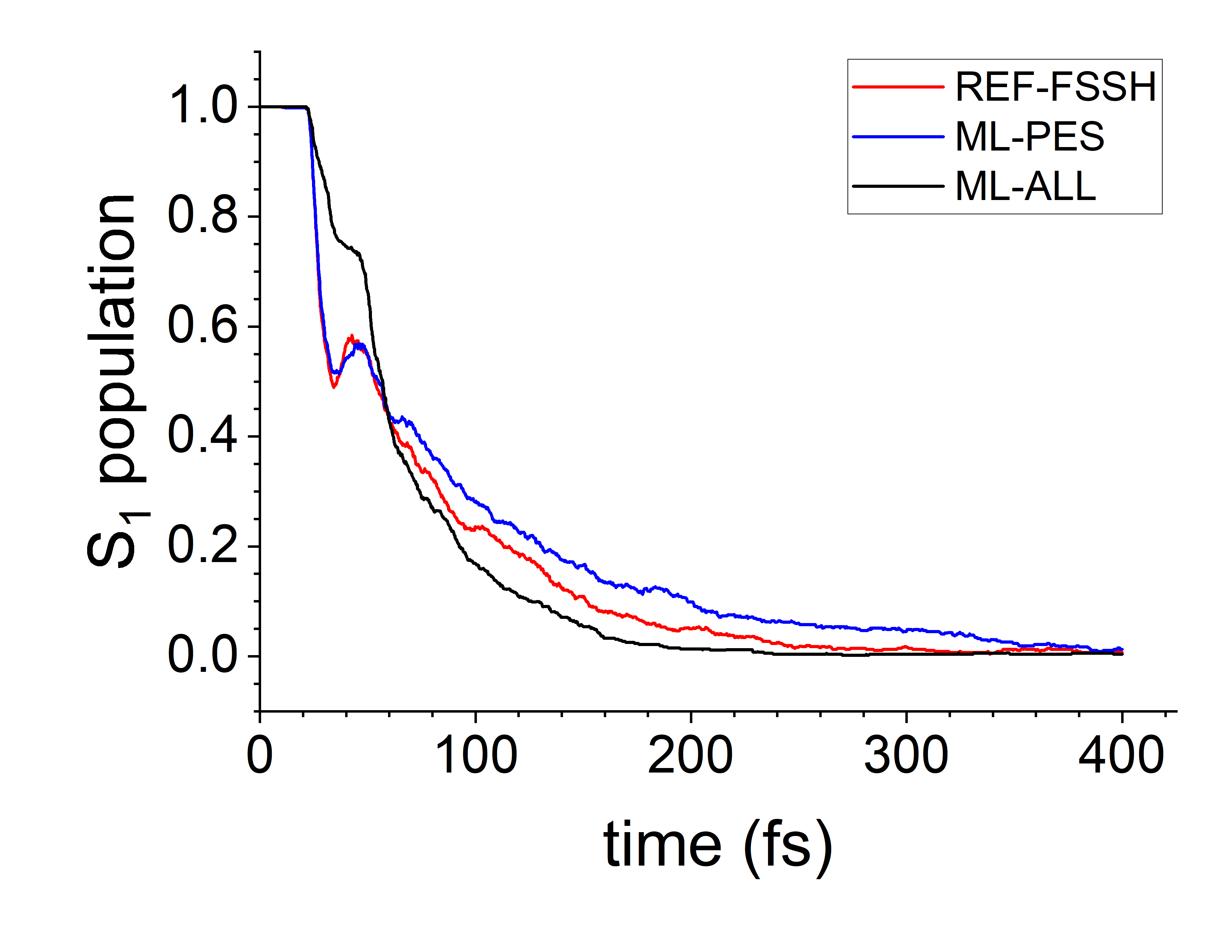}
        \caption{}
        \label{fig:excetiedlife4}
    \end{subfigure}
    \hfill
    \begin{subfigure}[b]{0.47\textwidth}
        \centering
        \includegraphics[width=\textwidth]{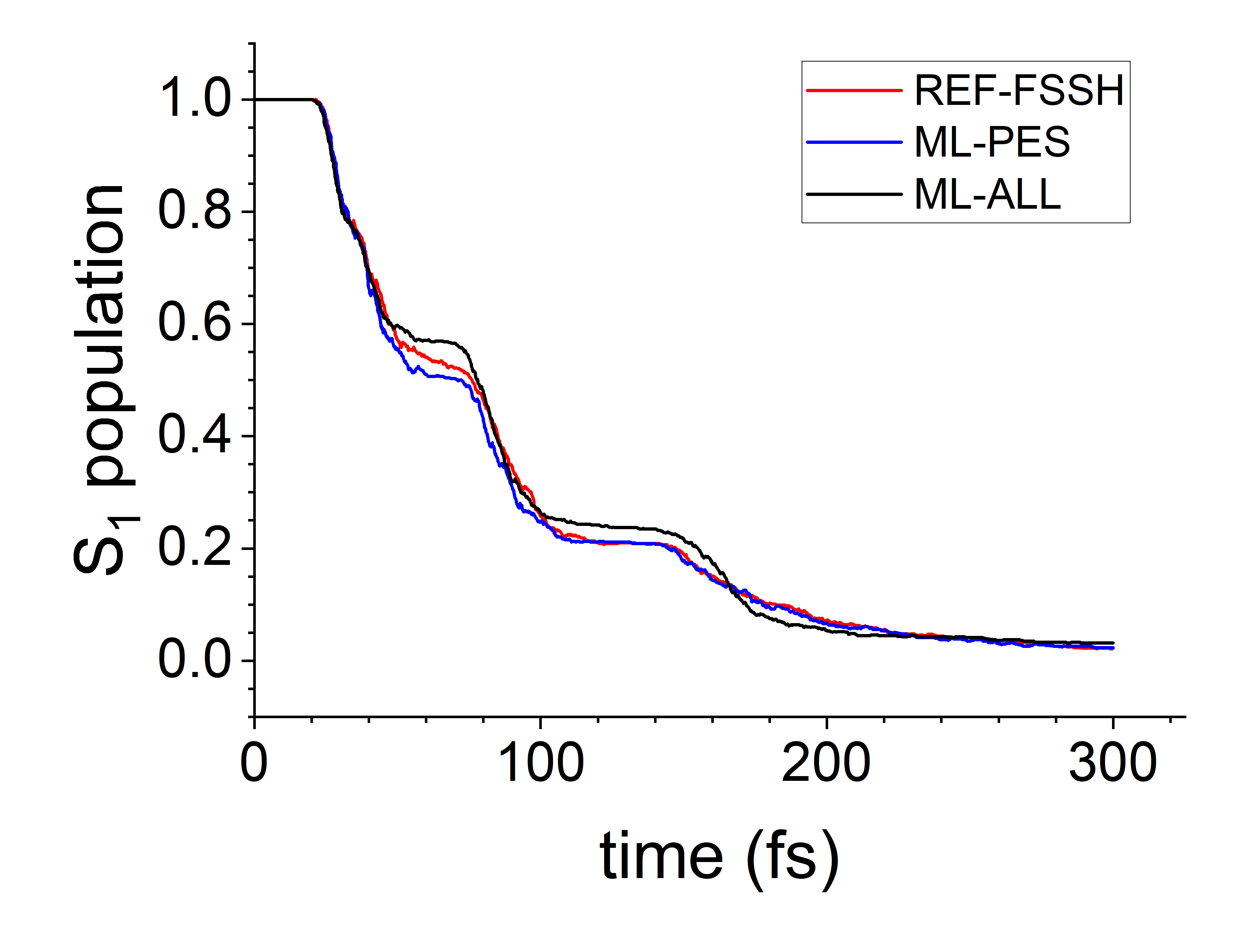}
        \caption{}
        \label{fig:Azolife4}
    \end{subfigure}
    \caption{Population of the electronic states $\mathrm{S_1}$ during photoisomerization of $\mathrm{CH_2NH}$ (a) and azobenzene (b). Different colors represent different methods: REF-FSSH in red, ML-PES in blue, and ML-ALL in black.}
    \label{fig:excetiedlife}
\end{figure*}

The time-dependent electronic state populations on the $\mathrm{S}_1$ state of $\mathrm{CH_2NH}$ can be seen in Figure \ref{fig:excetiedlife4}. The result of ML-PES coincides with FSSH in the first 60 fs but exhibits a slight deviation subsequently. It might be due to the mismatch between adiabatic PESs and nonadiabatic couplings. The former was predicted approximately using NequIP, while the latter was obtained precisely using external electronic structure calculation programs. If the regions with strong nonadiabatic coupling do not match the regions where the corresponding potential energy difference is allowed, the transition from $\mathrm{S}_1$ to $\mathrm{S}_0$ that should occur would be blocked, leading to an increase in the lifetime of the excited state (70.4 fs with REF-FSSH and 80.4 fs with ML-PES, see Table \ref{tab:Collectiveresults}). The $\mathrm{S}_1$ lifetime predicted with ML-ALL was 68.8 fs and very close to the reference value of 70.4 fs. However, error cancellation appears in the time-dependent $\mathrm{S}_1$ population curve. The underestimation of LSTM is relevant to its transition pattern, such as broader distribution of hopping events on nuclear degrees of freedom and more rapid dynamic behavior around conical intersection. More details can be seen in Section \ref{subsec:hopping locus}.

As shown in Figure \ref{fig:Azolife4}, both ML-PES and ML-ALL applied to azobenzene can reproduce the reference curve of its time-dependent $\mathrm{S}_1$ population. The corresponding simulations lead to the lifetime of 84.2 and 91.4 fs, respectively, which were in good agreement with the REF-FSSH prediction of 88.2 fs. Notably, the lifetime of excited azobenzene was slightly underestimated by ML-PES while overestimated by ML-ALL, which was opposite to $\mathrm{CH_2NH}$. The former might be due to different potential energy gaps in strong coupling regions, which would be further discussed in Section \ref{subsec:hopping locus}. The latter can be explained based on the representative test trajectory shown in Section \ref{subsec:Representative trajectory test}.  Notably, the first plateau in the $\mathrm{S}_1$ population curve may correspond to the sudden change of electron density matrix around 50 fs during simulations. It was missed out in the LSTM-driven test trajectory, making the system stay in the excited state longer.

Photochemical product yields of two reactions were estimated according to the molecular configurations in the last simulation step of total trajectories. The $cis$-to-$trans$ transformation of $\mathrm{CH_2NH}$ is determined by artificially marking the hydrogen atom connected to the nitrogen. The product yields were predicted as 0.55, 0.50 and 0.50 with REF-FSSH, ML-PES and ML-ALL, respectively. Note that a small deviation from 0.50 can be attributed to the occurrence of nitrene-like configurations. The photochemical product yields of $cis$-azobenzene were predicted as 0.46, 0.43 and 0.46 using REF-FSSH, ML-PES and ML-ALL, respectively. Our proposed ML method can accurately reproduce photoreaction yields of realistic molecules, at least for the present test examples.

\subsection{Distribution of hopping events}
\label{subsec:hopping locus}
\begin{figure*}[ht]
    \centering

    \subcaptionbox{\label{fig:hopping_locus_a}}{
        \includegraphics[width=0.30\textwidth, trim=30pt 8pt 35pt 8pt, clip]{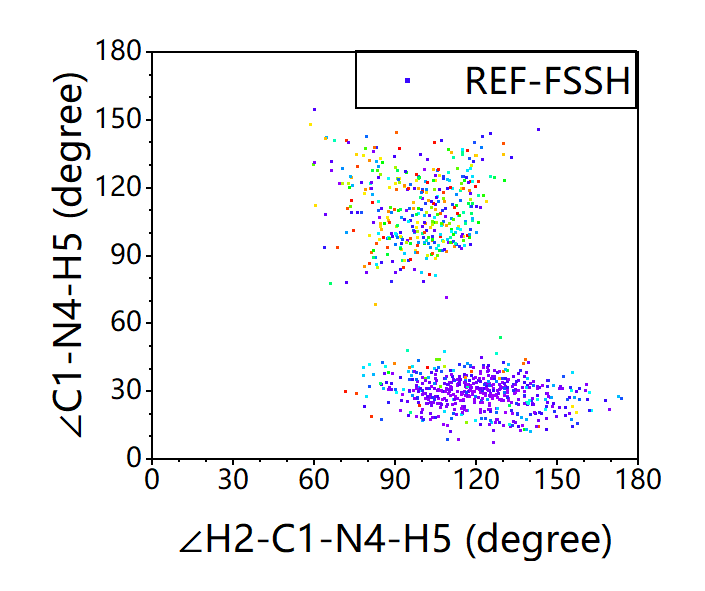}\hfill
        \includegraphics[width=0.30\textwidth, trim=30pt 8pt 35pt 8pt, clip]{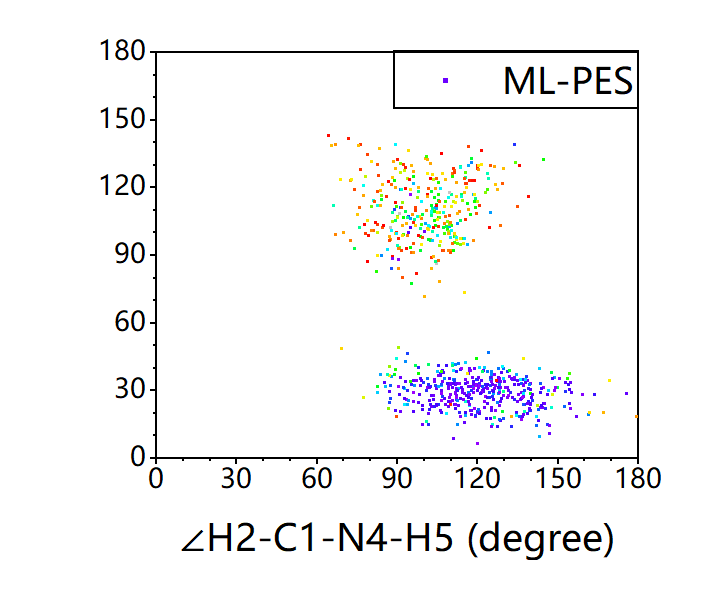}\hfill
        \includegraphics[width=0.36\textwidth, trim=30pt 4pt 20pt 20pt, clip]{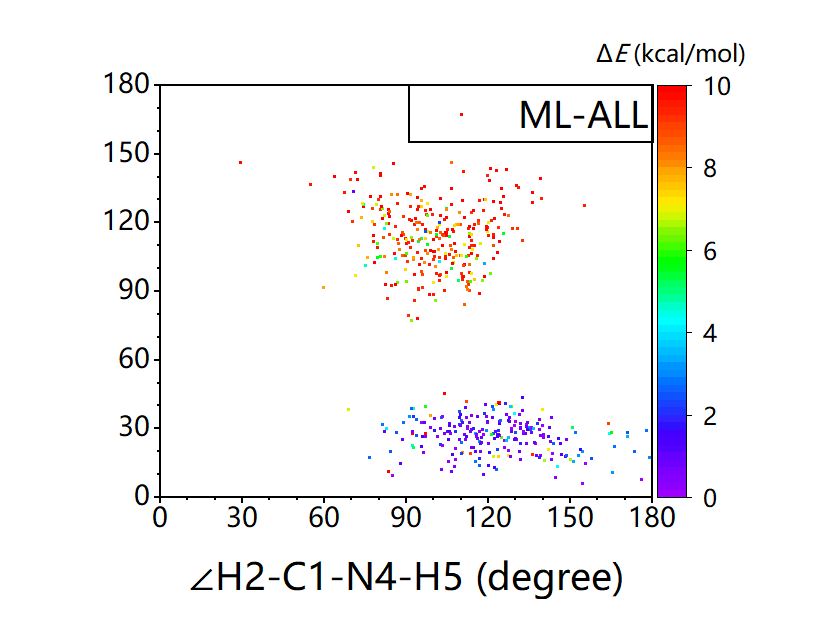}
    }
    \vspace{1ex}

    \subcaptionbox{\label{fig:hopping_locus_b}}{
        \includegraphics[width=0.30\textwidth, trim=40pt 8pt 35pt 8pt, clip]{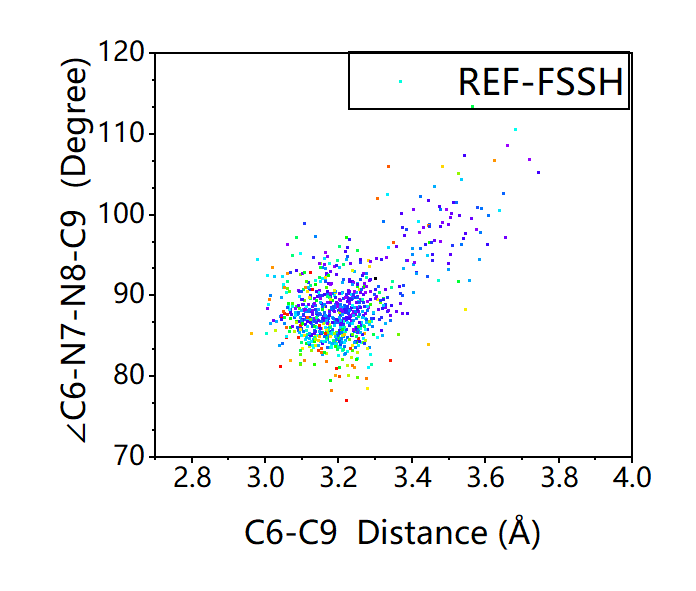}\hfill
        \includegraphics[width=0.30\textwidth, trim=40pt 8pt 35pt 8pt, clip]{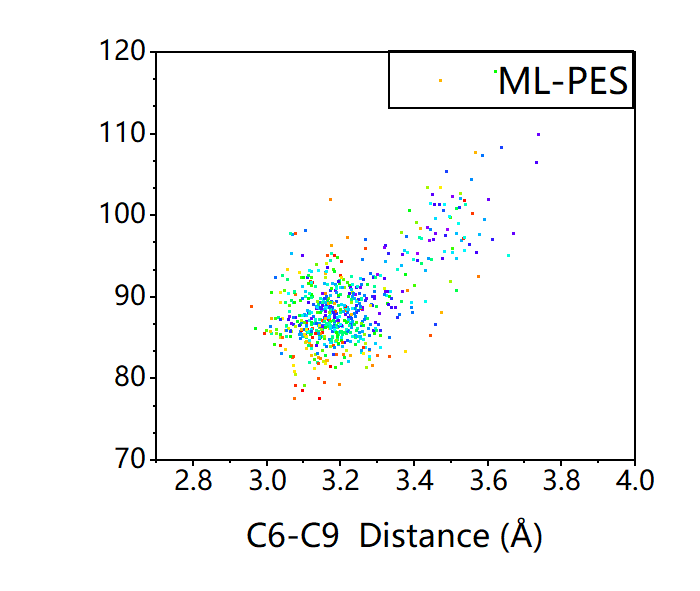}\hfill
        \includegraphics[width=0.36\textwidth, trim=40pt 20pt 40pt 20pt, clip]{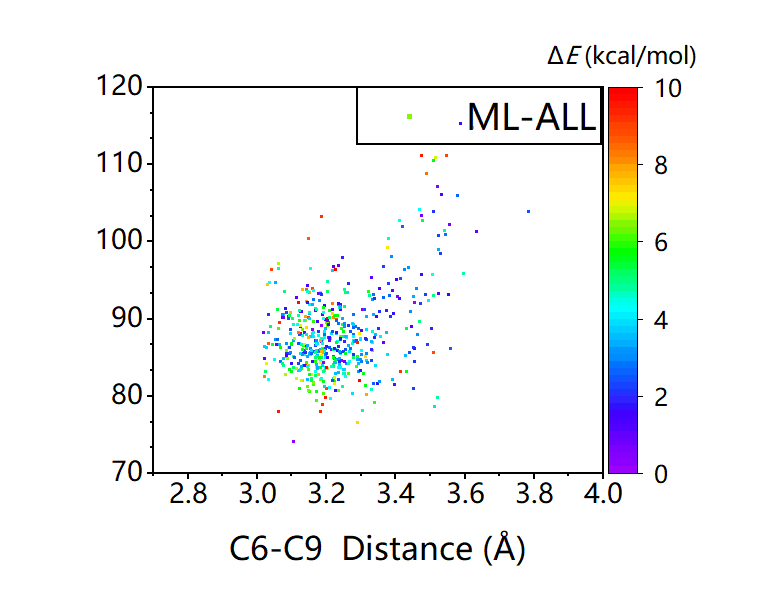}
    }

    \caption{Distributions of hopping events during photoisomerizations of $\mathrm{CH_2NH}$~(a) and azobenzene~(b) using REF-FSSH (left), ML-PES (middle), and ML-ALL (right). Different colors represent different values of $E_{\mathrm{S1}} - E_{\mathrm{S0}}$ when transition event occurs.}
    \label{fig:hopping locus}
\end{figure*}

To better demonstrate the reliability of our proposed ML framework, the distribution of hopping events from $\mathrm{S}_1$ to $\mathrm{S}_0$ was mapped to low-dimensional space. The results of different simulation methods were shown in Figure \ref{fig:hopping locus}. The density of dots indicates the probability of surface hopping, while the color represents the value of potential energy gap between $\mathrm{S}_0$ and $\mathrm{S}_1$. 

Observations on the hopping events during photoisomerization process of $\mathrm{CH_2NH}$ were in good agreement with the collective results. Firstly, we focused on the dots with larger $\angle \mathrm{C1}$-$\mathrm{N4}$-$\mathrm{H5}$. The energy gaps were increased in presence of ML, making an incongruence between the intersection of PESs and the region with strong nonadiabatic coupling. During ML-PES simulations, it finally led to less frequent transitions to the ground state as well as the overestimated excited-state lifetime. However, the influence on ML-ALL simulations is opposite. Unlike ML-PES in which electron density matrix evolves based on the calculated $d_{jk}$, ML-ALL disregarded nonadiabatic coupling vectors and predicted the change of $\rho_{00}$ directly, enabling surface hopping more rapidly once the system approaches the intersection region. It does not only led to the underestimation of lifetime with LSTM but also resulted in the broader distribution of hopping events. 

Additionally, the dots with smaller $\angle \mathrm{C1}$-$\mathrm{N4}$-$\mathrm{H5}$ in Figure \ref{fig:hopping_locus_a} correspond to nitrene-like configurations of $\mathrm{CH_2NH}$. These short-lived configurations appear infrequently in a single trajectory but can be sampled in many trajectories in the REF-FSSH dynamics. On one hand, the dynamic behavior of ML-PES in the region is similar to that of REF-FSSH, indicating good performance of NequIP in a broad configurational space. On the other hand, the training of LSTM becomes much more challenging in presence of these configurations, which may result in fewer hopping events in the region using ML-ALL. Nevertheless, the occurrence of nitrene-like configurations may lead to unstable dynamic trajectories, which would be excluded from collective results. The influence can be thus ignored in the present case.

The hopping events during photoisomerization of $cis$-azobenzene can be seen in Figure \ref{fig:hopping_locus_b}. Similar to Figure \ref{fig:hopping_locus_a}, the use of NequIP increased the potential energy gap in the coupling region. However, most of transitions took place when the energy gap was smaller than 5 kcal/mol. It is quite different from the dynamic behavior of $\mathrm{CH_2NH}$, in which a hopping attempt with an energy gap near 10 kcal/mol was often observed. As mentioned above, such transitions would be rejected during ML-PES simulations of $\mathrm{CH_2NH}$. In contrast, most of hopping attempts were accepted for azobenzene because of smaller energy gaps. Thus, the deviation of ML-predicted $\mathrm{S}_1$ lifetime of azobenzene can be neglected compared to that of $\mathrm{CH_2NH}$. 

As observed in Figure \ref{fig:hopping locus} and in our previous work, the performance of ML is prone to deteriorate around conical intersections. A weighted sampling related to potential energy difference should be useful when selecting training points from reference trajectories. However, whether the additional samplings are really effective remains to be studied on realistic molecules. It would be difficult to determine an appropriate weight in prior if multiple relaxation channels should be considered. Inclusion of rare configurations may seriously impede the training of LSTM models, such as the case of $\mathrm{CH_2NH}$ discussed above. We will take care of these issues in the future.

\section{CONCLUSIONS AND OUTLOOK}
\label{sec:CONCLUSIONSANDOUTLOOK}
In this work, we constructed a machine learning framework that consists of the NequIP potential energy surface prediction model and the LSTM electron density evolution prediction model. Starting from our previous work, the input features and modeling procedure of LSTMs were significantly modified for high-dimensional systems. It was successfully applied to fewest switches surface hopping, which is the most widely used nonadiabatic dynamic method, to simulate realistic molecules. Aside from NequIP for fitting adiabatic PESs, LSTM acts as the time propagator of electronic degrees of freedom, providing an alternative solution to deal with nonadiabatic couplings as the key to surface hopping. We employed two typical photochemical systems as $\mathrm{CH_2NH}$ and azobenzene, and validated that the proposed ML models can produce $\mathrm{S}_1$ lifetimes and product yields satisfyingly in comparison with conventional FSSH simulations. Despite the high dimensionality of molecular systems, only 10 reference trajectories were sufficient for LSTM training, followed by very efficient ML-driven dynamic simulations to generate trajectory ensemble. 

There is much room for improvement. First, the construction of excited-state PESs, even if without the need of fitting nonadiabatic coupling, is more challenging than that in the ground state. The success of ML-based PESs on excited $\mathrm{CH_2NH}$ has been demonstrated in recent years, while the complexity of ML algorithms is more essential to excited-state dynamic simulations on azobenzene as well as its derivatives. Although deep learning techniques such as SchNet and NequIP have exhibited their advantage according to our experience, the trade-off between computational accuracy and overhead remains a nontrivial issue, especially when dealing with larger molecules. OMNI-P2x\cite{Martyka2025,Martyka2025ANI} and other $\Delta$-learning\cite{10.1021/acs.jctc.5b00099} models may provide alternatives to our future work.
Moreover, constructing correct topology for conical intersections with machine learning remains an open challenge in the field, especially when the dataset is completely based on adiabatic trajectory snapshots. There are several machine learning and diabatization strategies can be considered.
\cite{Yarkony20_1848,Truhlar20_6456,Richardson23_011102,Schuurman23_7780,Gu23_6557}
The rate of trajectory failure, which is relatively higher than that in our previous works without ML, is also an issue to be solved. 
On one hand, training set consists entirely of physically plausible configurations sampled from reference trajectories, and trained model lacks the ability to identify edge conditions and cannot produce correct restoring forces to guide the system back to the physical manifold, leading to unphysical dissociation and bring bias. Teaching to model to recognize and penalize unphysical geometries would help. 
On the other hand, since the computational cost has been significantly reduced with the assistance of ML, a little more 'wasting' trajectory is not a problem in practice.

Second, the validation of LSTM models in step 4 is the bottleneck of the whole procedure. Similar to our previous work, here we still compared the time evolution of $\rho_{00}$ in representative test trajectories visually, which depends on our prior understanding of simulated systems. More rigorous and automated analysis method is necessary. Additionally, only a few LSTM models built in step 3 (e.g., 4\% for azobenzene) can exhibit superior performance, leading to a cumbersome iteration between checking and rebuilding. How to enhance the success rate of LSTM training is important to achieve higher efficiency.

Third, applications on more diverse systems are attractive but require modifications on the present framework. For example, the transition probability in eq \ref{eq:hopping judge two state} cannot deal with photochemical processes involving more than two electronic states. The global flux surface hopping\cite{GFSH10.1021/ct5003835} algorithm developed by Wang and coworkers offers a potential solution, but the growing size of electron density matrix would lead to more outputs of LSTMs and more complex network structures.
Correctly incorporating decoherence corrections will be another key objective. Future iterations of the LSTM-FSSH framework may require the integration of physics-informed layers or differentiable modules to ensure a robust ML propagator specifically account for the sudden damping effects introduced by decoherence algorithms.
Generalization of LSTM-driven surface hopping for simulating both internal conversion and intersystem crossing processes on an equal footing seems to be straightforward. However, intersystem crossing is always a rare event compared to internal conversion around conical intersections, posing a significant challenge for LSTM training.

Finally, our designed input features of LSTM can be replaced with an equivariant network in principle. Descriptor-free equivariant neural networks have achieved remarkable success across a wide range of molecular modeling tasks.\cite{MACE23_arXiv,Deshpande24_arXiv,Xu25_47044} Recent studies have demonstrated direct embedding of velocity vectors via spherical harmonics into equivariant architectures as input and output features.\cite{brandstetter2022geometricphysicalquantitiesimprove}
However, equivariant neural networks in current stage cannot process time-series data, and spatiotemporal equivariant neural network framework that simultaneously addresses both spatial symmetries and temporal evolution remains largely unexplored. 
Such an architecture would need to account for both the instantaneous equivariance of the molecular state and the cumulative spatiotemporal effects over time $t$. 
Our long-term goal is to construct a unified spatiotemporal equivariant network that facilitates time evolution based solely on position and velocity vectors (as well as initial conditions of electronic degrees of freedom), without relying on external descriptors, thereby performing nonadiabatic dynamics under even more rigorous physical backgrounds. The concept of spatiotemporal equivariance was only recently formally proposed,\cite{Keller25_arXiv} and future iterations of the LSTM-FSSH framework will likely be built upon these developments.

\section*{Supporting Information}
Introduction and details of NequIP, details of LSTM, negative examples for validation of representative trajectories, and supplemental results of two photochemical systems.

\section*{Conflict of Interest}
The authors declare no conflict of interest.

\section*{Data Availability}
The source code is available on \url{https://github.com/uminominami/LSTM-NequIP-code-v1}

\section*{Acknowledgements}
We are grateful to the financial support from the Natural Science Foundation of China (22573008, 22193041 and 22288201) and the Robotic AI-Scientist Platform of the Chinese Academy of Sciences.

\bibliography{v13.bib,tangdd}


\section*{Supplementary material}

# Table of Contents

# 1 Introduction and details of NequIP

In this work, we employed the Neural E(3)-equivariant Interatomic Potential (NequIP) to build potential energy surfaces (PESs) in ground and excited states for dynamics simulations. Conventional neural networks (NNs) for PES construction are typically designed to be invariant under rotations and translations, often relying on handcrafted descriptors such as interatomic distances. Although this ensures invariance of the final output, the internal representation cannot transform in a physically consistent way under symmetry operations, limiting its ability to capture directional dependencies and geometric structure. In contrast, not only the total potential energy but also the atomic forces predicted by equivariant neural networks are invariant under rotation and translation. If the atomic coordinates are rotated, the forces rotate accordingly.

Unlike many NN-based potentials that rely on rotationally invariant scalar descriptors such as interatomic distances or angles, NequIP operates directly on relative position vectors pointing from atom $i$ to atom $j$ as

$$\vec{r_{ij}} = \vec{r_i} - \vec{r_j} \tag{S1}$$

which preserves directional information and enables the network to capture anisotropic interactions. Such directional information is further mapped into spherical harmonic space, forming a complete basis for irreducible representations of the SO(3) group, and decomposing angular dependencies into well-defined angular momentum channels. The convolutional kernels in NequIP can be subsequently built as the product of a learnable radial function and the spherical harmonic basis function:

$$S_m^{(l)}(\vec{r}_{ij}) = R(r_{ij}) Y_m^{(l)}(\hat{r}_{ij}), \tag{S2}$$

where $r_{ij}$ and $\hat{r}_{ij}$ denote the length and unit vector of $\vec{r}_{ij}$, respectively, $Y_m^{(l)}(\hat{r}_{ij})$ is the spherical harmonic function, and $R(r_{ij})$ is the radial function modeled by a multilayer perceptron. This factorization separates radial and angular degrees of freedom explicitly, ensuring that the

filter responses transform covariantly under 3D rotations.

Based on the spherical harmonic filters, the features in NequIP are propagated through equivariant tensor products as the following layer:

$$V_{\text{out}}^{(l_o,p_o)} \propto \sum_{b\in\mathcal{N}(i)} \sum_{m_i,m_f} C^{l_o m_o}_{l_i m_i l_f m_f}\, Y^{(l_f)}_{m_f}(\hat{r}_{ib})\, R(r_{ib})\, V_b^{(l_i,p_i)}. \tag{S3}$$

Similar to the angular momentum theory in quantum mechanics, the Clebsch-Gordan (CG) coefficients serve here as the coupling constants that govern how irreducible representations of SO(3) combine under tensor products. The equivariance of this operation is obvious: the entire framework, which encodes directional information into spherical harmonics and combines angular channels via the CG coefficients, is built upon a mathematical formalism that inherently respects 3D rotational symmetry. This tensor product layer also provides a mechanism to dynamically adjust the feature dimensions. By selecting which output channels to include, the network can expand or compress the representation across layers. For example, high-rank tensor features with $l \geq 2$ can be gradually reduced to vectors ($l = 1$) or scalars ($l = 0$) in the last layer, enabling a symmetry-preserving dimensionality reduction.

After multiple layers of equivariant message passing as above, the high-dimensional tensor features $V_i^{(l,p)}$ associated with atom $i$ are gradually transformed and eventually projected down to the desired output space. In the NequIP for predicting potential energies, all features are reduced to scalar ($l = 0$) via an equivariant readout process. If only the invariant scalars are retained in the last layer, the total potential energy can be obtained by summing over these atomic contributions

$$E = \sum_i \alpha_i^{(0)} \cdot V_i^{(0)} \tag{S4}$$

where $\alpha_i^{(0)}$ denotes learnable weights. It is worth noting that while the total potential energy is a rotationally invariant scalar, the internal representation throughout the above network consists of non-trivial high-order geometric tensors, which transform covariantly under E(3) operations. Alternatively, some other equivariant NNs directly predict atomic forces as

vector outputs with $l = 1$, in which a rotation of the input structure induces a corresponding rotation of the force vectors.

Unlike conventional NNs that always discard angular information via scalarization at the beginning, NequIP preserves directional information and propagates it across layers, capturing complex angular dependencies with fewer training samples. By embedding physical symmetries directly into the network architecture, NequIP achieves superior generalization from limited data no more than a few thousand training configurations, and produces physically consistent predictions across diverse chemical environments from isolated molecules to extended periodic materials. For rigorous mathematical derivations and proofs of equivariant NNs, we refer the reader to the original papers about tensor field networks,[S1] 3D steerable convolutional neural networks,[S2] and Clebsch-Gordan Nets.[S3]

The NequIP models used in this work consist of 4 interaction blocks with irreducible representations up to $l_{\max} = 2$ of spherical harmonics for encoding. The cutoff distance for message passing was set to be 4.0 Å. The batch size and initial learning rate for optimization were 5 and 0.01, respectively, accompanied with early stopping based on the validation loss with a patience of 1000 epochs, the maximum allowed validation loss of $10^4$, and the minimum allowed learning rate of $10^{-6}$. The loss weights for potential energy and atomic forces were set as 1.0 and 0.1, respectively.

## 2  Details of LSTM

The structure of our proposed LSTM model was shown in Figure S1. As mentioned in the main text, the input features include nuclear coordinates, nuclear velocities, elements of electron density matrix, potential energies, and index of active state. Consider a two-state photochemical reaction with $N$ atoms. The size of nuclear coordinate features is $M \times N \times N_{SOAP}$, where $M$ denotes the length of previous time series extracted from dynamic trajectory, and $N_{SOAP}$ is determined by the hyperparameters of SOAP ($n_{max}$ and $l_{max}$). The size of nuclear velocity features is the same. The nuclear coordinate and velocity features are independently processed by an encoder, leading to a set of hidden variables with a reduced size of $M \times N_{nuc}$. The current encoder consists of the first dense layer with $N_{den1}$ nodes, a reshape layer, and the second dense layer with $N_{den2}$ nodes. Two sets of $M \times N_{nuc}$ hidden variables and the remaining input features are subsequently concatenated. Since two electronic states are involved in the present simulations, four elements of electron density matrix, two inputs derived from potential energies, and one index of active state should be considered. Thus, the size of the output of encoder is $M \times (2 \times N_{nuc} + 7)$. Then the variables are successively processed via the first attention layer with $N_{atten1}$ heads, five LSTM layers with $N_{LSTMx}$ ($x = 1$-5) nodes, the third dense layer with $N_{den3}$ nodes, the second attention layer with $N_{atten2}$ heads, a reshape layer and the fourth dense layer with $N_{den4}$ nodes, leading to the predicted elements of electron density matrix in the next time step. *tanh* was employed as the activation function of dense layers. The values of all hyperparameters used in this work were listed in Table S1.

Unlike conventional ML training procedures in which the prediction accuracy of test samples (i.e., a time series of density matrix in this work) should be examined carefully, reproduction of representative trajectories with the trained LSTM is much more important to validate ML performance. In this work, we paid less attention to the training procedure in order to reduce computational cost, especially for high-dimensional realistic molecular systems. In Step 3, we first performed several rounds (e.g., 4 for CH$_2$NH and 6 for azobenzene)

of LSTM training. After the preliminary optimization on LSTM parameters, we continued to train and saved the LSTM models every round. For each set of hyperparameters, 15 LSTM models with different parameters were saved. Finally, we picked one model with the smallest loss from them to validate whether the representative trajectories can be reproduced. On account of the fact that only a small portion of LSTMs are able to reproduce the reference trajectories, the above strategy is essential to enhance the efficiency of ML modelling. In contrast, all attempts on fine-tuning on optimizers were ineffective in practice. We applied the default Adam optimizer with batch size of 6 and learning rate of $10^{-3}$ in this work.

Table S1: Hyperparameters used for LSTM

| | Hyperparameter | Value | Description |
|---|---|---|---|
| **SOAP** | $r_{\text{cut}}$ | 4 | cutoff radius for local chemical environment in unit of Å |
| | $\sigma$ | 1.0 | variance of Gaussian function for atomic density in unit of Å |
| | $n_{\max}$ | 2 | number of radial basis functions |
| | $l_{\max}$ | 1 | maximum degree of spherical harmonics |
| | $N_{\text{SOAP}}$ | 42 | number of SOAP descriptors |
| **Encoder** | $N_{\text{den1}}$ | 3 | size of the first dense layer for encoding |
| | $N_{\text{den2}}$ | 15 | size of the second dense layer for encoding |
| | $N_{\text{nuc}}$ | 15 | dimensionality of nuclear DoFs after encoding ($= N_{\text{den2}}$) |
| **LSTM** | $N_{\text{atten1}}$ | 4 | number of heads in the first attention block |
| | $N_{\text{LSTMx}}$ | 96 | size of the $x$-th LSTM layer, $x = 1 \sim 5$ |
| | $N_{\text{den3}}$ | 4 | size of the third dense layer after LSTM processing |
| | $N_{\text{atten2}}$ | 4 | number of heads in the second attention block |
| | $N_{\text{den4}}$ | 4 | size of the last dense layer ($=$ dimensionality of electronic DoFs) |
| | $M$ | 32 | number of frames in a sequence (time series) for training |
| | $M_{\text{shift}}$ | 2 | sequence shift for extracting the next sequence |

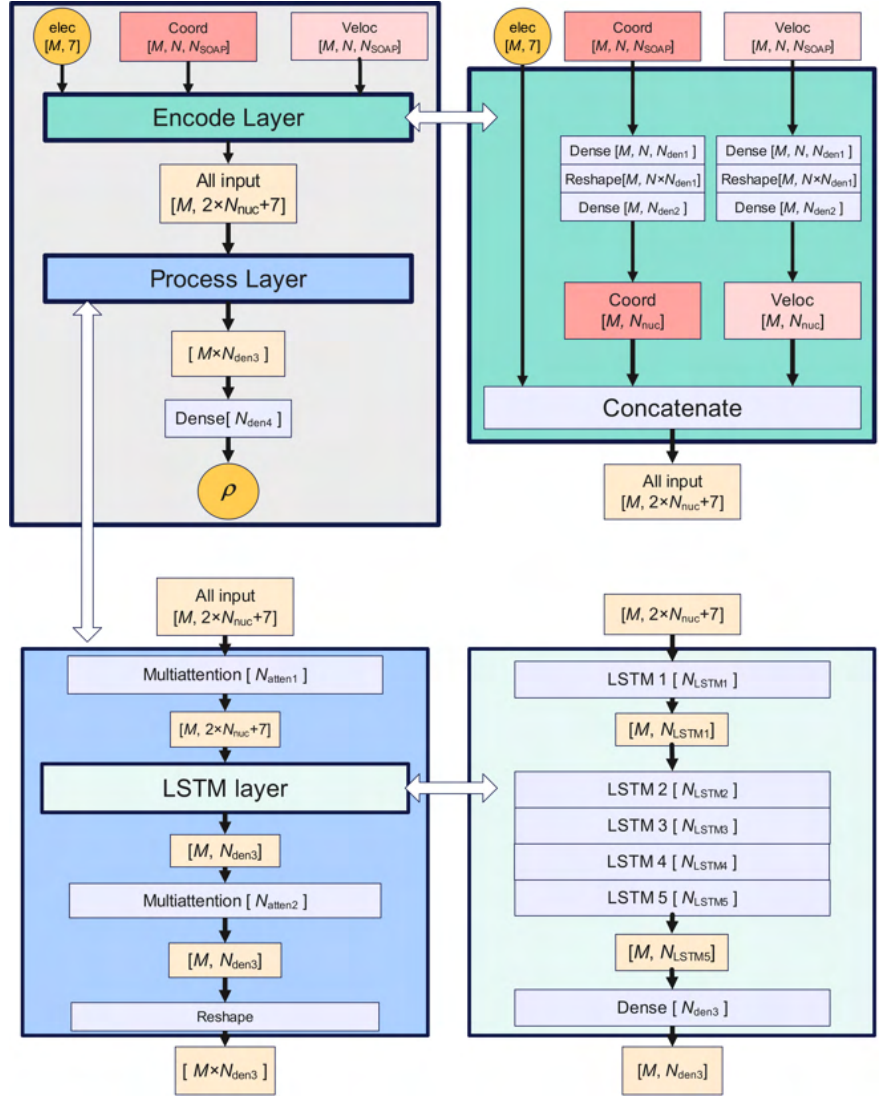

Figure S1: Detailed structure of LSTM networks. $N$ denotes the number of atoms. $M$ denotes the number of frames in a sequence (i.e., sequence length).

# 3 Negative examples for validation of representative trajectories

As discussed in the main text, only a small portion of LSTMs are able to reproduce the representative trajectory. In absence of a quantitative criterion to measure the similarity of two FSSH trajectories, we have no choice but to perform such comparisons visually and laboriously.

For two-state nonadiabatic processes, we focus on the time evolution of $\rho_{00}$. On account of the correction of eq 20 in the main text, the prediction accuracy of off-diagonal elements of density matrix mainly depends on the performance of $\rho_{00}$. Fortunately, most of negative results of $\rho_{00}$ are easy to identify. Some negative examples of $CH_2NH$ were listed in Figure S2. which can be excluded as follows:

(1) The trivial trajectory without any change of $\rho_{00}$ throughout dynamic simulations should be excluded firstly, such as (b) and (l).

(2) The trajectory that jumps to other electronic state at the very beginning of simulations should be excluded, such as (c), (d), (e), (f), (g), (h), (j), (k) and (o). Obviously, these trajectories would deviate from the reference throughout dynamic simulations.

(3) The trajectory missing the first hopping event indicates overestimation of excited-state lifetime, such as (a) and (n). The relevant LSTM is slightly better than those in (1) and (2). If the LSTM-driven trajectory exhibits similar pattern to the reference in the latter stage of dynamic simulation, the LSTM can be employed as the initial model for further optimization.

(4) The trajectory that reproduces the first hopping event but exhibits a different pattern is better than those in (1), (2) and (3). However, we observed that the final collective results obtained using the corresponding LSTMs are still deviated from the reference values. For example, the oscillation after 50 ps in the reference trajectory is completely lost in (i) but overestimated in (m). Further optimization starting from these LSTMs is strongly suggested.

Negative examples of azobenzene were shown in Figure S3. Two thirds of them stay in the $S_1$ state throughout dynamic simulations. No transition takes place in these trajectories, making a serious mistake compared to the reference. The remaining trajectories capture the first hopping to the ground state at 20 fs, including (a), (b), (l) and (n). The re-hopping event back to the excited state, which is observed immediately in the reference trajectory, vanishes in (l) and appears too late in (b); (a) and (b) also undergo too frequent transitions between two states before 100 fs. Among these five trajectories, (a), (b) and (l) finally relax to the ground state as the same as the reference. None of them leads to satisfactory collective results during the following LSTM-driven dynamic simulations. Only (a) performs slightly better than others.

According to our experience, hundreds of attempts are necessary to obtain the best LSTM model with satisfactory reproducibility (as shown in Figure 5 in the main text). The use of more sophisticated NNs for PES construction can increase the successful rate, especially for medium-size or larger molecules. Fortunately, the training and validation processes have been modified in this work, reducing the computational cost significantly. As a result, so many visual comparisons are tedious for researchers but affordable for computational resources.

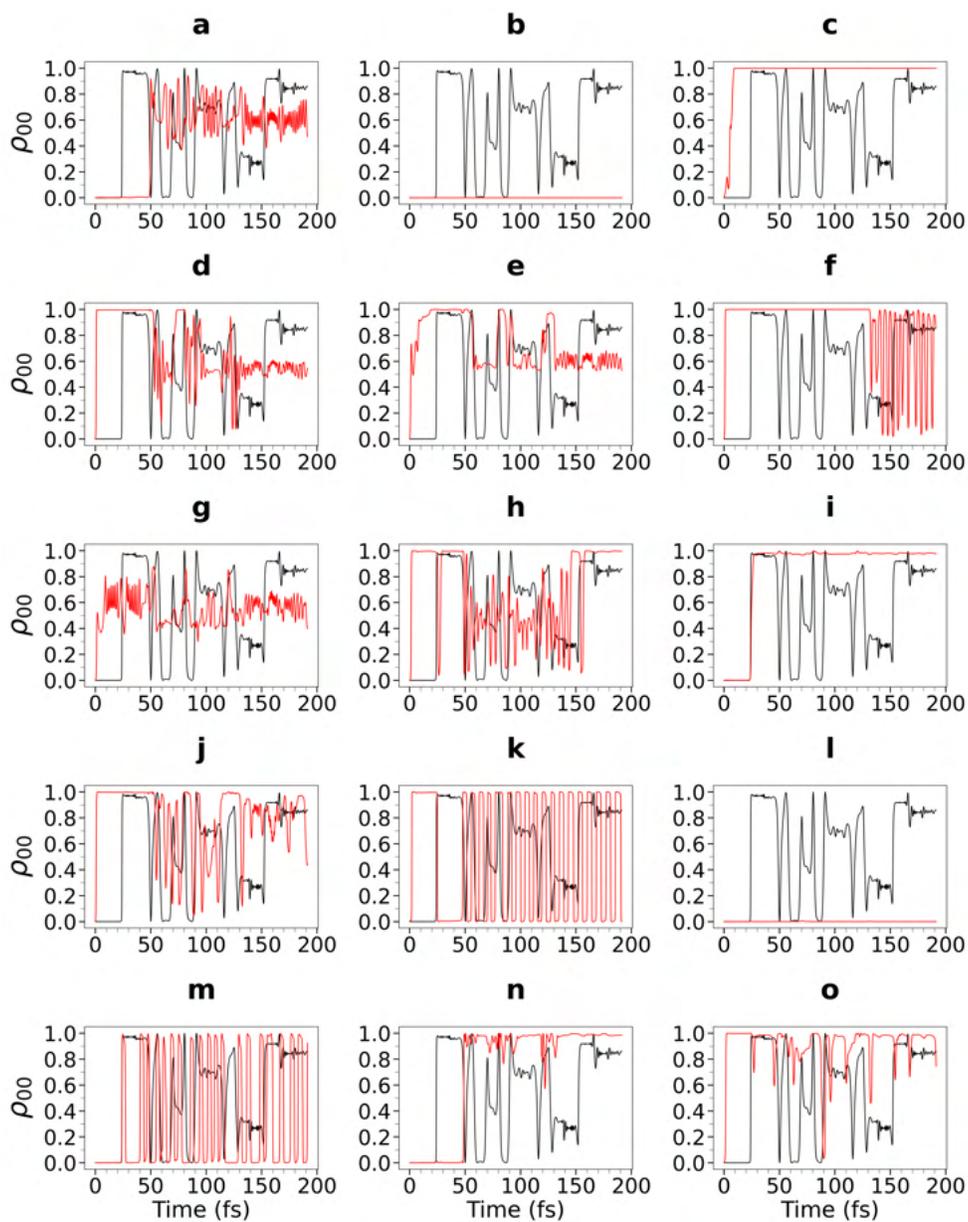

Figure S2: Negative examples for validation of a representative trajectory of $CH_2NH$. Different colors represent different methods: reference FSSH in black, LSTM-FSSH in red.

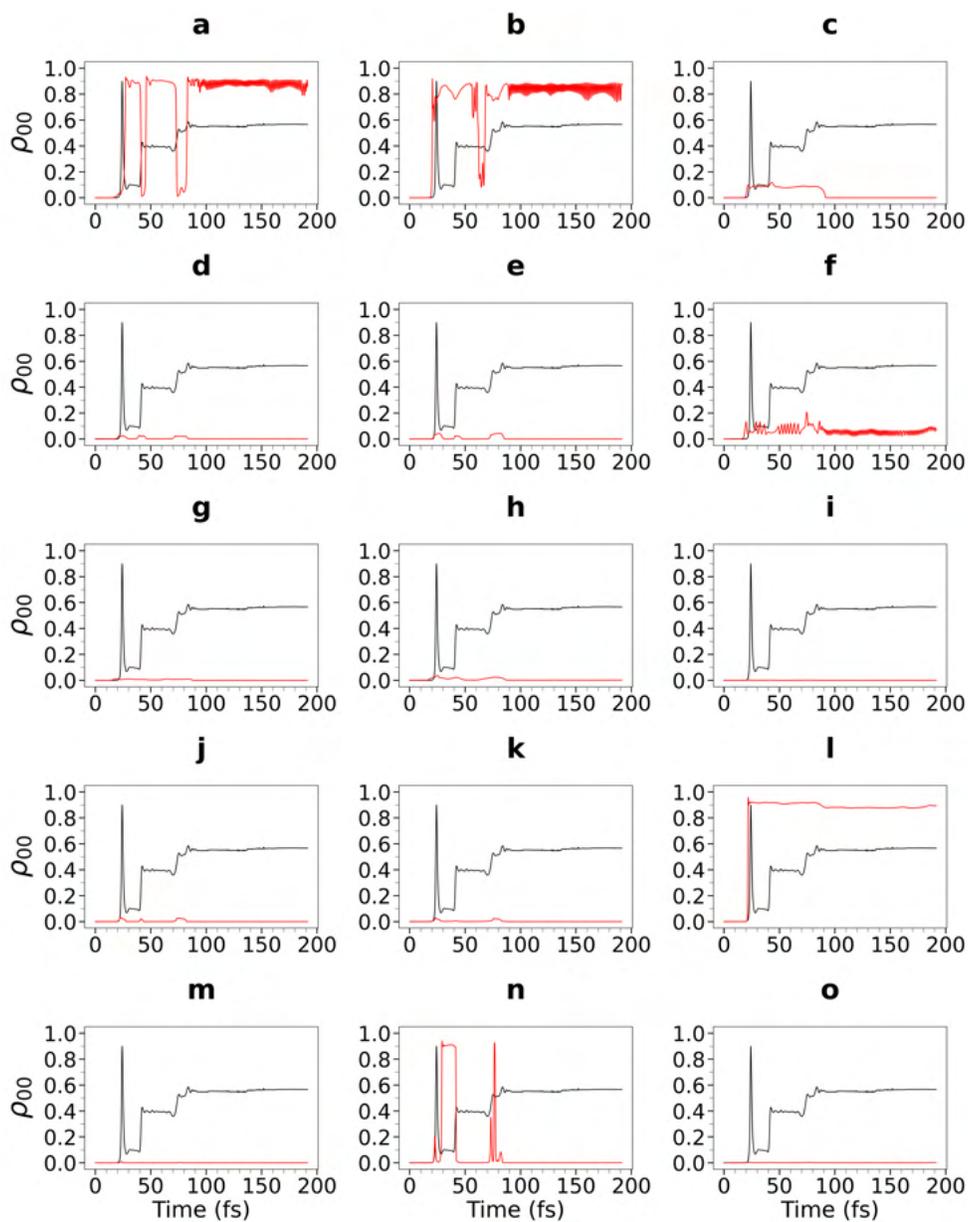

Figure S3: Negative examples for validation of a representative trajectory of azobenzene. Different colors represent different methods: reference FSSH in black, LSTM-FSSH in red.

# 4 Supplemental Results

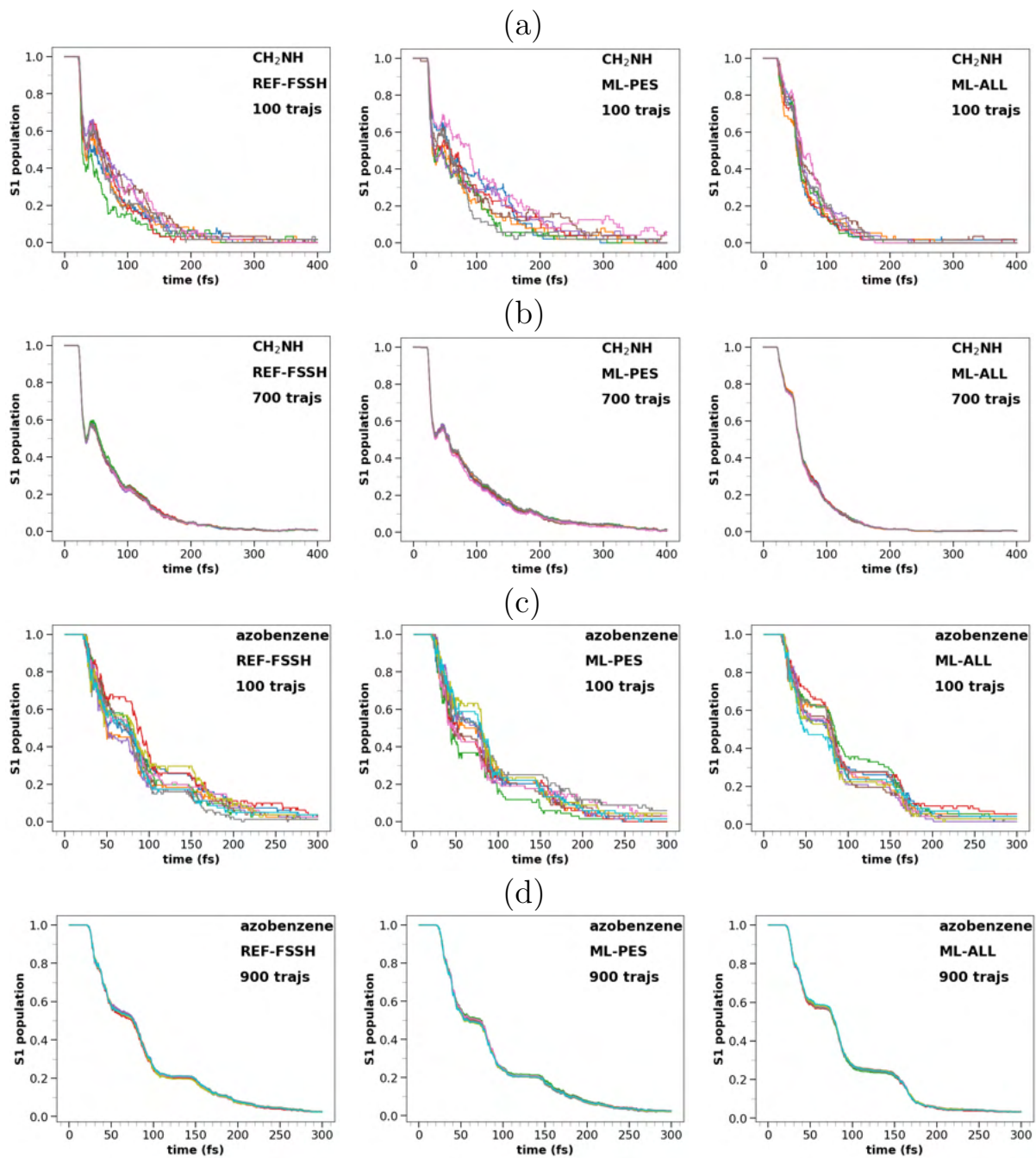

Figure S4: Convergence of trajectory ensembles. For FSSH simulations of $CH_2NH$, when 100 trajectories are selected from 800, the time-dependent $S_1$ state populations are divergent (a). In contrast, when 700 are selected from 800, the result achieves good convergence (b). Similarly, for simulations of azobenzene, when 100 trajectories are selected from 1000, the time-dependent $S_1$ state populations are divergent (c). In contrast, when 900 are selected from 1000, the result achieves good convergence (d).

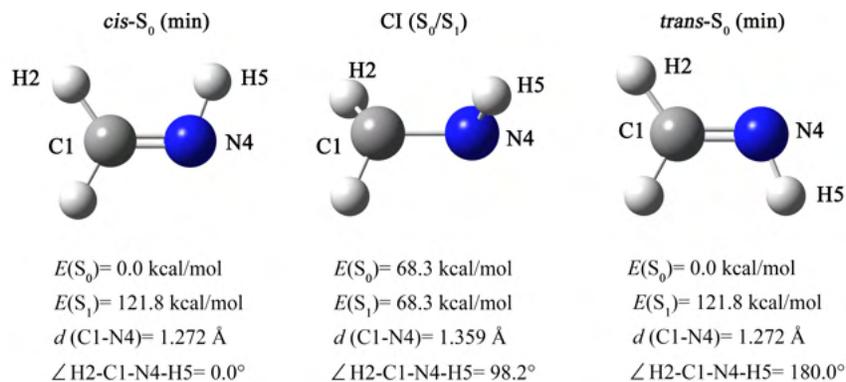

(a)

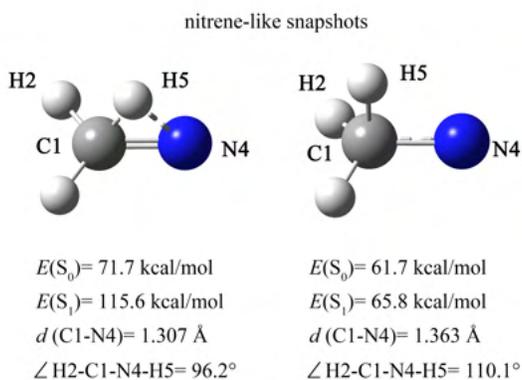

(b)

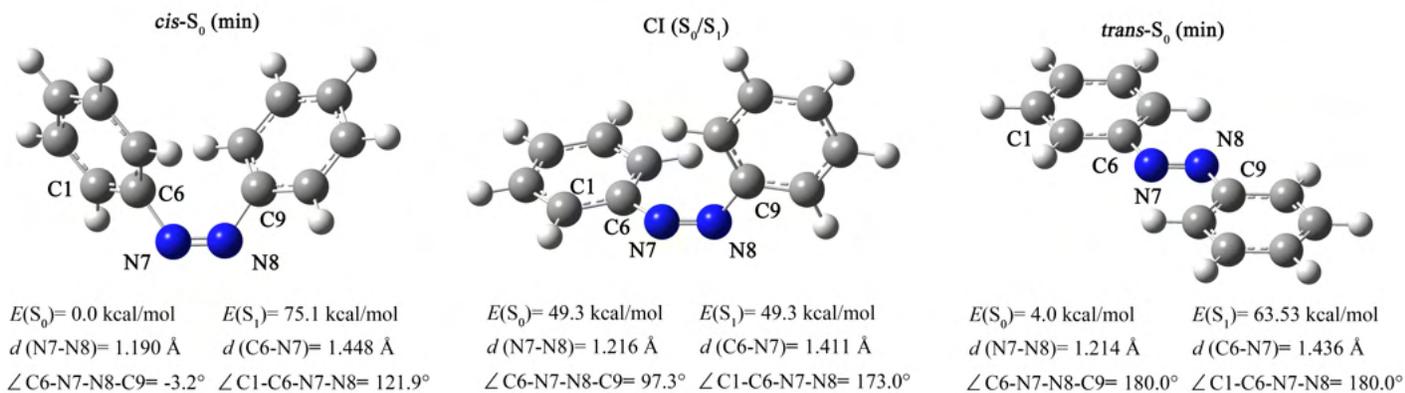

(c)

Figure S5: Key configurations with potential energies of two test systems: minima and conical intersection of CH$_2$NH optimized at the CASSCF(2,2)/6-31G(d) level (a), representative nitrene-like configurations of CH$_2$NH extracted from trajectories (b), minima and conical intersection of azobenzene optimized at the OM2/MRCI level (c).

 

\newpage
\begin{tocentry}
\centering
\includegraphics[width=3.25in]{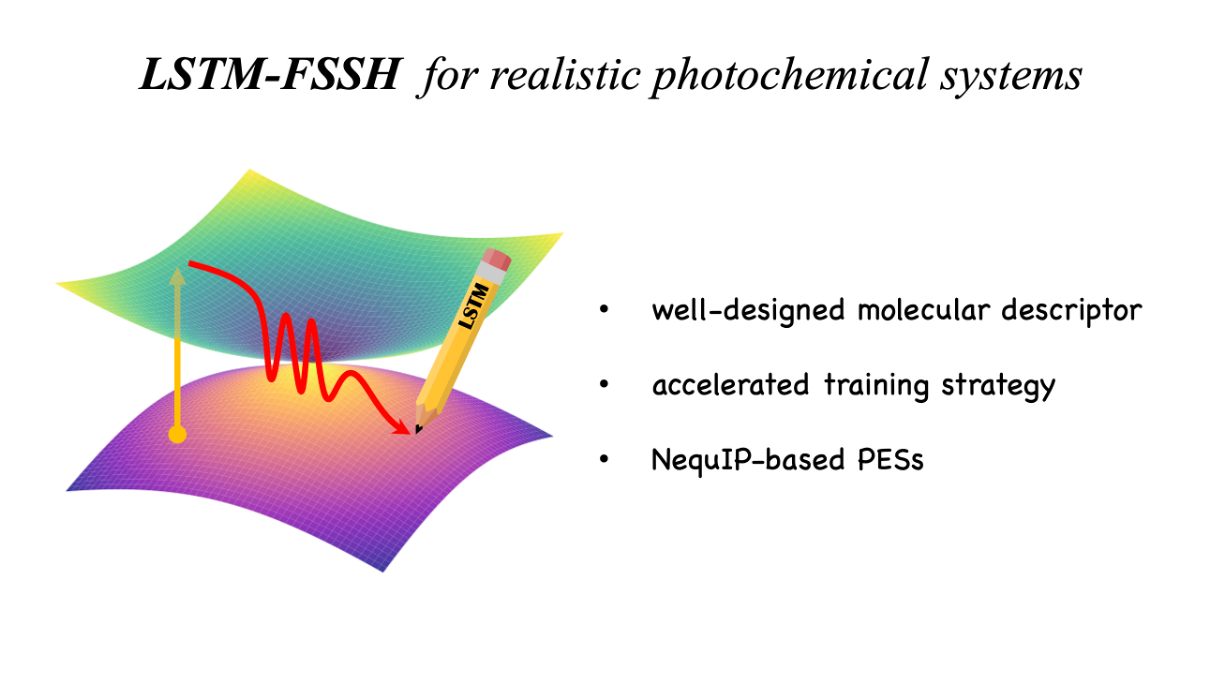}
\end{tocentry}

\end{document}